\documentclass[sn-mathphys]{sn-jnl}

\jyear{2021}%

\theoremstyle{thmstyleone}%
\theoremstyle{thmstyletwo}%

\theoremstyle{thmstylethree}%

\begin{document}

\title[Complexity factor of spherically anisotropic polytropes...]{Complexity factor of spherically anisotropic polytropes from gravitational decoupling.}


\author[1]{\fnm{C.} \sur{Las Heras}}\email{camilo.lasheras@ua.cl}
\equalcont{These authors contributed equally to this work.}

\author*[1]{\fnm{P.} \sur{Le\'on}}\email{pablo.leon@ua.cl}

\affil*[1]{\orgdiv{Departamento de F\'isica}, \orgname{Universidad de Antofagasta}, \orgaddress{\street{Aptdo 02800}, \city{Antofagasta}, \country{Chile}}}


\abstract{In this work we will analyse the complexity factor, proposed by L. Herrera, for spherically symmetric static matter distributions satisfying a polytropic equation through the gravitational decoupling method. Specifically, we will use the 2-step GD, which is a particular case of the extended geometric deformation (EGD), to obtain analytic polytropic solutions of Einstein's equations. In order to give an example, we construct a model satisfying a polytropic equation of state using Tolman IV as a seed solution.}




\maketitle

\section{Introduction}

The theory of polytropes has been useful to describe realistic stellar models. Polytropic equations of state have been widely used in astrophysics since they allow one to deal with a large variety of physical phenomena \cite{Chandrasekhar,Schwarzschild,shapiro2,kippenhahn,Kovetz,Goldreich,Abramowicz,Tooper,Bludman,Nilsson,Herrera9,Lai,Thirukkanesh,Shojai,zdenek1}. The general formalism to study polytropes for anisotropic distributions was presented in \cite{Herrera12} and \cite{Herrera13,Herrera14}, for the Newtonian and general relativistic regimes, respectively. For relativistic matter distribution, a possible polytropic equation of state (and the one we are interested in in this work) is given by
\begin{eqnarray}
P_r &=& K \rho^{1+1/n},
\end{eqnarray}
where $P_r$ is the radial pressure, $\rho$ the energy density, $K$ the polytropic constant and $n$ the polytropic index. The polytropic constant can be used to describe two different scenarios. The first one is when $K$ can be written in terms of other physical constants. This case has been used to describe white dwarfs and leads to the Chandrasekhar mass limit \cite{kippenhahn}. The second case is when $K$ is a free parameter, which is useful to describe completely convective stars \cite{kippenhahn}. It is also related to Bose-Einstein condensates with repulsive ($K>0$) or attractive ($K<0$) self-interaction \cite{Chavanis}.

For a spherically symmetric isotropic matter fluid, the polytropic equation of state is enough to solve Einstein’s equations. However, when local anisotropy in pressure is considered, it is necessary to provide further information about the internal structure of the system. One common choice to close the system of equations is to propose a particular anisotropy profile by giving the anisotropic function (see \cite{Herrera13,Herrera14,Abellan2,Abellan3,Contreras17}). Another possibility is to choose a particular form of the energy density (see, for example \cite{Nunez2}). Nevertheless, it is known that the latter leads to a singular tangential sound velocity at the surface of the distribution for $n>1$ \cite{Nunez2}. One solution to this problem is to consider a generalized polytropic state equation, given by
\begin{eqnarray}\label{polytropicEoS}
P_r &=& K \rho^{1+1/n}+ G \rho + D,
\end{eqnarray}
where $G$ and $D$ are constants. For recent developments or applications see \cite{Nunez,Stuchlik,Stuchlik2,Stuchlik3,Stuchlik4} and references therein. The constant $G$ may be used to describe different scenarios. Some of the then include: radiation ($G=1/3$), pressureless matter ($G=0$ if $K=0$) and vacuum energy ($G=-1$). The polytropic component dominates the linear component for high densities when $n>0$ and for low densities when $n<0$ \cite{Chavanis2}. On the other hand, the constant $D$ allows to consider polytropic models with a discontinuous energy density at the surface of the distribution

Now, there are several reasons to expect local anisotropy in the pressures of self-gravitating objects \cite{Bowers}. Deviatons of isotropy or fluctuations in the anisotropy can be produced by a wide range of physical phenomena that may occur in the interior of matter distributions as compact objects. See for example \cite{Herrera10,Herrera2,Herrera11,Herrera3} and references therein. Some examples are the intense magnetic fields observed in neutron stars and white dwarfs, the high viscosity produced by neutrino trapping or the exotic phase transitions that may occur in the gravitational collapse, among others. The impact of the anisotropy can be studied with Newtonian gravity for densities lower than $10^{14} g/(cm)^3$. However, general relativistic effects have to be considered for higher densities, as in the case of neutron stars or white dwarfs. The equation of state to describe such a  system must also be anisotropic. Some examples of anisotropic relativistic stars in GR and beyond can be found in \cite{Rincon5,Rincon6,Rincon7,Rincon8} and references therein.

Another interesting feature of the anisotropy function of stellar distributions is that it determines the complexity factor of the distribution together with the gradient of the energy density, according to the definition proposed in \cite{Herrera6}, for spherically symmetric static fluid distributions. Although it is not directly related to the concept of entropy or disequilibrium, the author suggests such a link might exist. However, this definition of the complexity in the context of the Einstein theory of gravitation it may be understood as a generalization of previous definitions which were completely determined by the energy density \cite{Lopez3,Sanudo,Chatzisavvas,DEAVELLAR,DEAVELLAR2,DEAVELLAR3,DEAVELLAR4}, in the sense that it allow us to take into account other physical factors that play a key role in the internal structure of self-gravitating systems. It is, by construction, intrinsically related to the internal structure of the system, and it is constructed by assuming the homogeneous perfect fluid distribution is one of the simplest systems that could be studied. This is, it has vanishing complexity. Later on, it was extended to dynamically spherically symmetric fluids \cite{Herrera7}, distributions with axial symmetry \cite{Herrera8} and vacuum solutions represented by the Bondi metric \cite{Herrera9}. See also \cite{Abbas,Sharif10,Sharif:2022wnl,Sharif:2021gsl,Zubair50,Zubair51,Yousaf,Yousaf2,Yousaf3,Yousaf4,Yousaf5,Yousaf40,Maurya:2022cyv,Maurya:2022cyv2,Contreras:2022vec2} for more recent developments. 

Now, a very useful and powerful procedure to search for solutions to Einstein’s equations is the known gravitational decoupling method \cite{Ovalle}. This was originally introduced in the context of Randall-Sundrum brane world \cite{Ovalle1,Ovalle15,Ovalle2,Ovalle16,Ovalle8,Ovalle9,Ovalle7,Ovalle10,Ovalle11} and later on used in the general relativity framework where the gravitational decoupling was proved. Specifically, this method allows one to study, in a very simple and systematic way, a self-gravitating system whose Einstein-Hilbert action is given by
\begin{eqnarray}
S = \int \left[\frac{R}{2k^2}+\mathcal{L}\right]\sqrt{-g}d^4x + \alpha \mbox{(corrections)},
\end{eqnarray}
which leads to energy-momentum tensor of the form 
\begin{equation}
    T_{\mu \nu} = T^0_{\mu \nu} +  \alpha \theta_{\mu \nu}.
\end{equation}
The source $\theta_{\mu \nu}$, can be interpreted as the coupling with other fluid distributions \cite{Our,Estrada1,Gabbanelli,Morales1,Morales2,Tello2,Contreras7}, as the coupling with others fields \cite{Ovalle13} or as contributions coming from theories beyond GR \cite{Sharif3,Sharif4,Sharif5,Tello4,Estrada3,Leon,Leon2}. 

Originally, the method only considered spatial metric deformations in spherically symmetric systems. It was named minimal geometric deformation (MGD). Its generalization also considers deformation of the temporal component, together with the spatial component of the metric, and it was named as extended geometric deformation (EGD) \cite{Ovalle12}. Moreover, it was already formulated for axially symmetric matter distributions in  \cite{Contreras13}. (For a list of additional applications of the gravitational decoupling method, see \cite{Ovalle6,Ovalle17,Ovalle18,Ovalle19,Ovalle20,Abellan,Abellan4,Sharif,Sharif2,Sharif8,Sharif9,Cavalcanti,Darocha1,Darocha2,Darocha3,Darocha4,Darocha5,Darocha6,Darocha7,Darocha8,Darocha9,Casadio2,Contreras,Contreras2,Contreras9,Contreras5,Contreras14,Contreras15,Rincon,Rincon2,Rincon3,Rincon4,Tello6,Tello7,Hensh,Maurya2,Maurya3,Maurya4,Maurya5,Zubair}).

The geometric deformation approach considers solutions to Einstein equations with a line element written in Schwarzschild-like coordinates. However, there are known exact solutions written in other coordinate systems whose transformation to Schwarzschild coordinates is not always well defined (see for example, isotropic coordinates \cite{Nariai}). In \cite{Our2}, we propose two inequivalent MGD-inspired algorithms that allow us to obtain new analytical and anisotropic solutions of Einstein equations in isotropic coordinates. It is important to mention that, although these two methods are based on the same metric deformations as MGD, they are not a rewriting of the MGD method in other coordinates. In fact, one of the methods does not lead to decoupling of the sources, while the other requires an additional condition (not considered in standard MGD) to achieve this. In \cite{Our3}, we proposed a new interpretation of EGD where the temporal and radial deformations are not simultaneous but consecutive. It leads to a set of solutions contained in EGD by solving simpler systems of equations. This method was named 2-step GD, and it is also valid to generalize the algorithms presented in \cite{Our2} in isotropic coordinates.

The use of polytropic equations of state within the context of geometrical deformation has been subtle. In \cite{Contreras4} is explored the inverse problem in the context of polytropic black hole on $(A)dS$ and in \cite{Contreras16} the polytropic equation of state has been used to construct a regular black hole in three dimensional gravity. More recently, it has been used in the context of $5D$ Gauss-Bonnet gravity \cite{Maurya7}. During the realization of this work, in \cite{Contreras20} the authors proposed a method to understand the effect of polytropic fluids on any other gravitational source by means of geometric deformation. Regarding the complexity factor within the framework of MGD, see \cite{Contreras8,Contreras18,Contreras19}, and more recently it has been studied in the extended case (EGD) in \cite{Maurya6}.

In this work, we will analyze the complexity factor of different polytropic solutions of Einstein’s equations using the extended version of the gravitational decoupling method. This is the geometric deformation considering both, temporal and spatial metric components. In particular, we will use the 2-step GD method to construct models satisfying a generalized polytropic state equation (\ref{polytropicEoS}). The complexity factor will then be evaluated in terms of $n$, $K$, and $G$. 

This paper is organized as follows: In Section 2, we present the basic set of equations that will be used in the other sections, that is, the Einstein's field equations for a spherically symmetric anisotropic fluid distribution and the complexity factor. In section 3, we summarize the extended gravitational decoupling method. In section 4, we present the 2-step GD method discussed in \cite{Our3}, and that will be implemented in the next section. In Section 5, we select Tolman IV as the seed solution in order to construct models satisfying a polytropic state equation using the left path of 2-step GD. We analyse the complexity factor of these families of solutions. We perform a detailed analysis of the physical acceptability conditions. In Section 6, we discuss the polytropic solution obtained, and in Section 7, we analyze how it can be applied to other polytropic models in terms of the baryonic mass density. In Section 8, we discuss all the results.

\section{Basics equations}
\subsection{The Einstein equations}

Let us consider a static, spherically symmetric distribution of an anisotropic fluid bounded by a surface $\Sigma$. In Schwarzschild-like coordinates, the metric is given by
\begin{eqnarray}\label{metric}
ds^{2}=e^{\nu}dt^{2}-\frac{1}{\tilde{\mu}}dr^{2}-r^{2}(
d\theta^{2}+\sin^{2}(\theta)d\phi^{2}),
\end{eqnarray}
where $\nu$ and $\widetilde{\mu}$ are functions of $r$ and must satisfy the Einstein equations
\begin{eqnarray}
    R_{\mu \nu} - \frac{1}{2}Rg_{\mu\nu} = \kappa^2 T_{\mu \nu},
\end{eqnarray}
where $R_{\mu \nu},R,T_{\mu \nu}$ are the Ricci tensor, the curvature scalar and the energy-momentum tensor, respectively. For the metric (\ref{metric}) the Einstein equations lead to the following system:
\begin{eqnarray}
\hspace{-0.5cm} \kappa^2 T^0_0&=& \frac{1}{r^{2}}-\frac{\tilde{\mu}}{r^{2}}-\frac{\tilde{\mu}'}{r} ,\label{ee1}\\
\hspace{-0.5cm} \kappa^2 T^1_1&=&\frac{1}{r^{2}}-\tilde{\mu}\left(
\frac{1}{r^{2}}+\frac{\nu'}{r}\right),\label{ee2} \\
\hspace{-0.5cm} \kappa^2 T^2_2 &=& - \frac{\tilde{\mu}}{4}
\left(2\nu'' +\nu'^{2}+2\frac{\nu'}{r}
\right) - \frac{\tilde{\mu}'}{4}\left(\nu'+\frac{2}{r}\right) \label{ee3},
\end{eqnarray}
where primes denote derivatives with respect to $r$. 

From the conservation law
\begin{eqnarray}\label{Dtmunu}
\nabla_{\mu}T^{\mu\nu}=0,
\end{eqnarray}
the equilibrium equation for anisotropic matter can be obtained
\begin{eqnarray}\label{TOV}
 -(T^1_1)'+\frac{\nu'}{2}(T^0_0 -T^1_1)-\frac{2}{r}\Delta=0,
\end{eqnarray}
where $\Delta= T^1_1-T^2_2$. 

Outside the fluid distribution, we shall assume that the space-time is given by the Schwarzschild exterior solution \cite{Schwarzschild2}, namely
\begin{eqnarray}
ds^{2}&=&\left(1-\frac{2M}{r}\right)dt^{2}-\left(1-\frac{2M}{r}\right)^{-1}dr^{2} - r^{2}(d\theta^{2}+\sin^{2}(\theta) d\phi^{2}).
\end{eqnarray}
Therefore, the continuity of the first and the second fundamental form across the boundary surface $r_{\Sigma}= constant$ implies that
\begin{eqnarray}
e^{\nu_{\Sigma}}&=&1-\frac{2M}{r_{\Sigma}},\label{nursig}\\
\mu_{\Sigma}&=&1-\frac{2M}{r_{\Sigma}},\label{lamrsig}\\
P_{r_{\Sigma}}&=&0, \label{psig}
\end{eqnarray}
where the subscript $\Sigma$ indicates that the quantity is evaluated at the boundary surface $\Sigma$.

\subsection{Complexity Factor}
For a spherically symmetric static fluid distribution, the complexity of the system is completely determined by the absolute value of the scalar function $Y_{TF}$ given by
\begin{eqnarray}
Y_{TF} = -\kappa \Delta - \frac{4\pi}{r^3}\int_0^r \tilde{r}^3 (T^0_0)' d\tilde{r}. \label{ComplexFactor_gen}
\end{eqnarray}
From this expression, it is clear that the complexity of the system is entirely characterized in terms of the energy density gradient and the anisotropic function. Thus, one of the simplest distributions is the homogeneous isotropic fluid, since each term is identically zero. However, this is not the only case with zero complexity. There is also the case in which the anisotropic function satisfies
\begin{eqnarray}
\Delta &=& -\frac{1}{2r^3}\int_0^r \tilde{r}^3 (T^0_0)' d\tilde{r} = \frac{r}{2\kappa^2}\left(\frac{\tilde{\mu}-1}{r^2}\right)' =  \frac{1}{\kappa^2}\left(\frac{3m}{r^3}-\frac{k^2}{2}T^0_0\right),
\end{eqnarray}
where
\begin{eqnarray}
    m =  \frac{\kappa^2}{2} \int^r_0 T^0_0 \bar{r}^2 d\bar{r}.
\end{eqnarray}
Let us emphasize that according to the original definition in \cite{Herrera6}, the first term on (\ref{ComplexFactor_gen}) corresponds to $-\kappa\Delta = \kappa\Pi$ with $\Pi=-\Delta = P_r-P_t$. Thus, we have a different sign in the first term compared to the expression used in other works (see for example \cite{Ovalle19,Maurya6}). This difference leads to different differential equations when imposing conditions on the complexity factor. This will be carefully discussed in \cite{Our4}.

Now, it can be shown that the complexity factor is directly related to the Tolman mass \cite{Herrera6} 
\begin{eqnarray}
    m_{T} = \frac{\kappa^2}{2} \int_0^{r} \bar{r}^2e^{(\nu + \lambda)/2}(T^0_0-T^1_1-2T^2_2)d\bar{r},
\end{eqnarray}
which in terms of $Y_{TF}$, can be written as
\begin{eqnarray}
m_{T} = (m_T)_{\Sigma}\left(\frac{r}{r_{\Sigma}}\right)^3 + r^3 \int^{r_{\Sigma}}_r \frac{e^{(\nu + \lambda)/2}}{\tilde{r}}Y_{TF} d\tilde{r}.
\end{eqnarray}
The first term in this equation is the Tolman mass of a homogeneous and isotropic fluid sphere of radius $r_{\Sigma}$. Therefore, the second term can be interpreted as the deviation of the Tolman mass from the homogeneous and isotropic fluid when both the anisotropy function and the energy density gradient are different from zero. However, as we mentioned before, the homogeneous and isotropic fluid is not the only case which leads to a vanishing complexity factor.

\section{Gravitational decoupling}
In this section, we shall summarize the gravitational decoupling method presented in \cite{Ovalle12}. The starting point of this method is to assume that the energy-momentum tensor can be written as
\begin{equation}
\label{3.1}
T_{\mu\nu}=T^{\rm 0}_{\mu\nu}+\alpha\,\theta_{\mu\nu},
\end{equation}
where $\alpha$ is a coupling constant. In this work we will assume that $T^{\rm 0}_{\mu\nu}$ is the matter-energy content associated with an anisotropic fluid. This is
 \begin{eqnarray}
 \label{3.2}
T^0_{\mu\nu}=(\rho+P_{\perp})u_{\mu}u_{\nu}-P_{\perp}g_{\mu\nu}+(P_{r}-P_{\perp})s_{\mu}s_{\nu},
\end{eqnarray}
where
\begin{eqnarray}
u^{\mu}=(e^{-\nu/2},0,0,0),
\end{eqnarray}
is the four velocity of the fluid and $s^{\mu}$ is defined as
\begin{eqnarray}
s^{\mu}=(0,\tilde{\mu},0,0),
\end{eqnarray}
such that $s^{\mu}u_{\mu}=0$, $s^{\mu}s_{\mu}=-1$.

The next step is to assume that the contribution of $\theta_{\mu\nu}$ to the complete system is encoded in the temporal and radial metric deformations, $h$ and $f$, respectively
\begin{eqnarray}
\label{3.4}
\nu
&=&
\xi+\alpha\,h
\ ,
\\
\label{3.5}
\tilde{\mu}
&=&
\mu+\alpha\,f
\ .
\end{eqnarray}
In this case, it is easy to check that, using (\ref{3.1}), (\ref{3.4}) and (\ref{3.5}), Einstein's equations (\ref{ee1})-(\ref{ee3}) are split into two systems. The first one coincides with Einstein's equation system of an anisotropic fluid
\begin{eqnarray}
\label{3.6} 
\hspace{-0.8cm} 8\pi(T^0)^0_0 & = & \frac{1}{r^2} -\frac{\mu}{r^2} -\frac{\mu'}{r}, \\
\label{3.7}
\hspace{-0.8cm} 8\pi (T^0)^1_1 & = & \frac 1{r^2}-\mu\left( \frac 1{r^2}+\frac{\xi'}r\right), \\
\label{3.8}
\hspace{-0.8cm} 8\pi (T^0)^2_2 & = & -\frac{\mu}{4}\left(2\xi''+\xi'^2+\frac{2\xi'}{r}\right) - \frac{\mu'}{4}\left(\xi'+\frac{2}{r}\right),
\end{eqnarray}
where $\kappa=8\pi$ and the corresponding conservation equation is given by
\begin{eqnarray}\label{TOV2}
 - [(T^0)^1_1]'+\frac{\nu'}{2}[(T^0)^0_0 -(T^0)^1_1]-\frac{2}{r}\Delta_0=0,
\end{eqnarray}
which is the TOV equation for an anisotropic fluid with $\Delta_0= (T^0)^1_1-(T^0)^2_2 $. 
The second system of equations reads
\begin{eqnarray}
\label{3.10}
 8\pi\,\theta_0^{\,0}
&\!\!=\!\!&
-\frac{f}{r^2}
-\frac{f'}{r}
\ ,
\\
\label{3.11}
8\pi\,\theta_1^{\,1} + Z_1
&\!\!=\!\!&
-f\left(\frac{1}{r^2}+\frac{\nu'}{r}\right)
\ ,
\\
\label{3.12} 8\pi\,\theta_2^{\,2} + Z_2
&\!\!=\!\!&
-\frac{f}{4}\left(2\nu''+\nu'^2+2\frac{\nu'}{r}\right) -\frac{f'}{4}\left(\nu'+\frac{2}{r}\right),
\end{eqnarray}
with 
\begin{eqnarray}
Z_1 & = & \frac{\mu h'}{r}, \\
4Z_2 & = & \mu \left(2h'' + \alpha h'^2 + \frac{2h'}{r}+2\xi'h'\right)+\mu'h'.
\end{eqnarray}
In order to find a solution of Einstein's equations for an energy-momentum tensor of the form (\ref{3.1}), we have to solve the systems (\ref{3.6})-(\ref{3.8})  and (\ref{3.10})-(\ref{3.12}). Now, if we assume that the set $\{T^0_{\mu\nu},\mu,\xi\}$ is a known solution of Einstein's equations, then it is only necessary to solve the second system. In both cases, there are more unknown functions than equations, so additional information is required in order to solve the system. 

It can be seen that the energy-momentum given by (\ref{3.1}), satisfies a conservation equation equivalent to (\ref{TOV}), written as a linear combination of equations (\ref{ee1})-(\ref{ee3}) with (\ref{3.1}), (\ref{3.4}) and (\ref{3.5}). It is important to note that the sources $T^{0}_{\mu \nu}$ and $\theta_{\mu \nu}$ can only be decoupled if there is an energy exchange between them. This can be easily seen from the conservation equations
\begin{equation}
\nabla_\mu (T^0)^{\mu}_\nu = -\frac{h'}{2}(P_r+\rho)\delta^1_\nu, \label{Cons1}
\end{equation} 
and 
\begin{equation}
\nabla_\mu \theta^\mu_\nu = \frac{h'}{2}(P_r+\rho)\delta^1_\nu, \label{Cons2}
\end{equation}
where the divergence in these expressions is calculated with the metric related to (\ref{3.4}) and (\ref{3.5}).

At this point, it is clear that EGD is a powerful tool to study more complicated solutions of Einstein's field equations than the ones obtained with the MGD method. Nevertheless, finding solutions to equations  (\ref{3.10})-(\ref{3.12}) could be very complicated depending on the system under study.
 
Finally, it is easy to show that the complexity factor associated with the energy-momentum tensor $T_{\mu \nu}= T^0_{\mu \nu}+ \alpha\theta_{\mu \nu}$ can be written as
\begin{eqnarray}
    Y_{TF} = Y^0_{TF} + \alpha Y^{\theta}_{TF}, \label{CF_EGD}
\end{eqnarray}
where $Y^0_{TF}$ is the complexity factor of the seed solution and 
\begin{eqnarray}
    Y^{\theta}_{TF} = -8\pi (\theta^1_1-\theta^2_2) - \frac{4\pi}{r^3}\int_0^r \tilde{r}^3 (\theta^0_0)' d\tilde{r},  \label{CF_Source}
\end{eqnarray} 
is the complexity factor of the source $\theta_{\mu\nu}$. 

\section{The 2-step GD}
\label{2-steps}
In this section, we shall briefly review the results presented in \cite{Our3} for the 2-step geometric deformation. This method is a simplification of the extended case discussed in the previous section. The idea is to find solutions to Einstein’s equations with both, temporal ($h$) and spatial ($f$) metric deformations, by performing non-simultaneous metric deformations. The gravitational decoupling method is used twice, once to deform only the spatial (temporal) metric component and once to deform only the temporal (spatial) metric component. Now, using the same notation as \cite{Our3}, we will call the left path (right path) when we perform a spatial (temporal) metric deformation followed by a temporal (spatial) metric deformation. 

\begin{figure}
\centering
\resizebox{7.5cm}{5.5cm}{
  \includegraphics{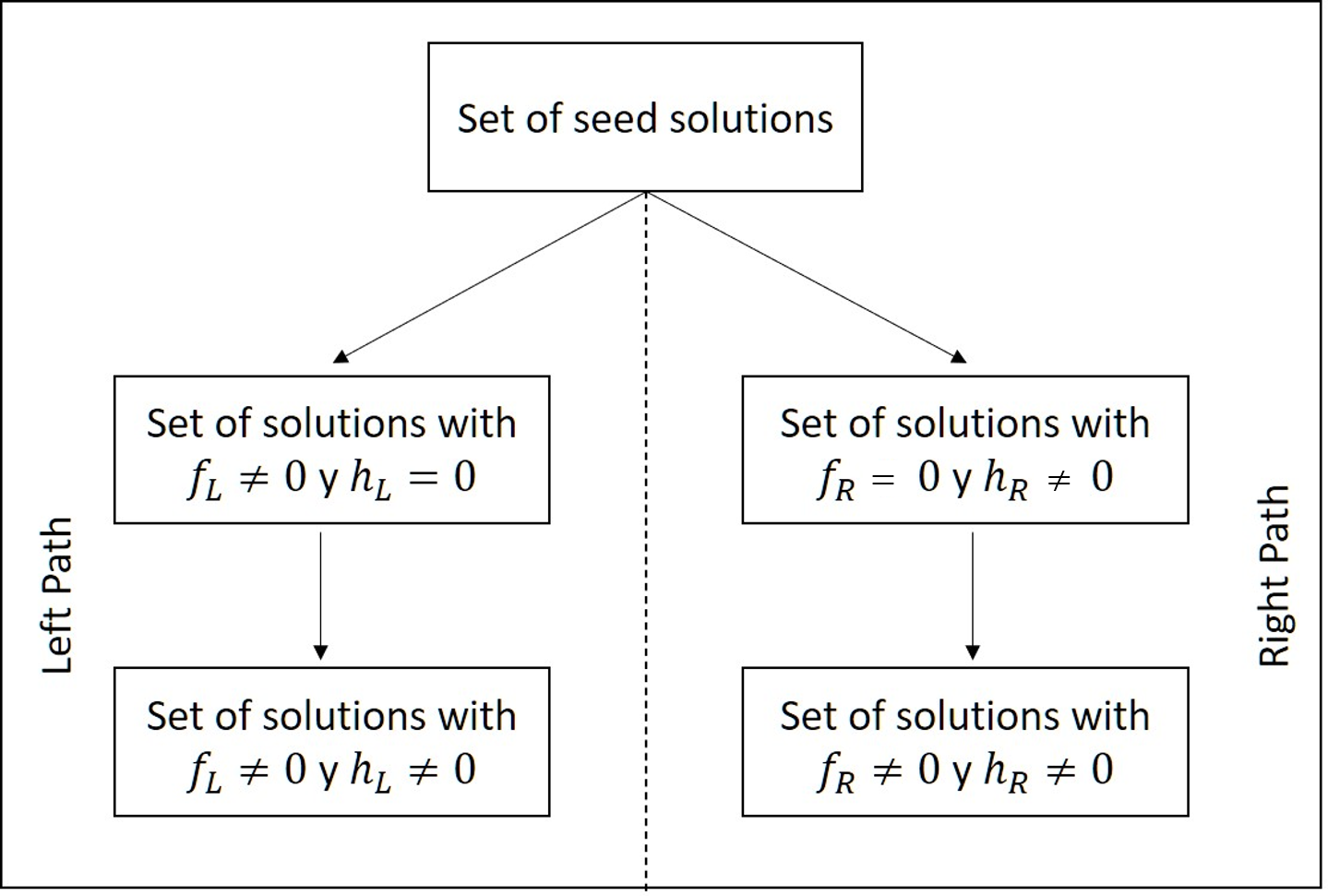}
}
\caption{The 2-step GD diagram}
\label{fig:1}      
\end{figure}

It should be noted that this procedure is a restriction of the extended geometric deformation. 2-step GD, in particular, necessitates a decomposition of the source $\theta_{\mu\nu}$ given by
\begin{eqnarray}
    \theta_{\mu \nu} &=&  (\theta^f)_{\mu \nu} + \frac{\beta}{\alpha} (\theta^h)_{\mu \nu},
\end{eqnarray}
where $(\theta^f)_{\mu \nu}$ and $(\theta^h)_{\mu \nu}$ are in charge of deformations of the metric's spatial and temporal components, respectively (See figure \ref{fig:1}). This particular decomposition of the source $\theta_{\mu\nu}$ is not required for EGD. Thus, there are solutions that could be obtained by the EGD but not by the 2-step GD. 

Let us emphasize that the $r$-dependence of the temporal deformation $h$ ensures that it is not a simple modification of the time scale. Indeed, it can be seen from the Ricci invariants that solutions related by temporal deformations are, in general, inequivalent (see appendix B of \cite{Our3}). This is because it will be related to a matter distribution with null energy density but with radial and tangential pressures different from zero. Therefore, is better to interpret the pure temporal metric deformations as mathematical artifacts to obtain new solutions to Einstein's equations (see for example \cite{Boonserm}). Thus, only the total energy momentum will have a physical interpretation. It is worth mentioning that \cite{Our3} was the first work to consider a purely temporal deformation in its scheme in order to obtain analytical solutions of Einstein equations with anisotropy in the pressures, that may describe realistic objects. More recently, it has been used to obtain solutions with a zero complexity factor \cite{Andrade}.

In the following, we shall write the basic systems of equations for the left and right paths of the 2-step GD, assuming that the seed solution is characterized by the set of functions $\{\xi,\mu,T^0_{\mu \nu}\}$.

\subsection{Left path}
The left path first considers the gravitational decoupling method with a pure spatial deformation. This is equivalent to applying the MGD, considering the solution characterized by $\{\xi,\mu,T^0_{\mu \nu}\}$ as seed solution. Here, the system that we need to solve is given by
\begin{eqnarray}
\label{Tolmanf1}
8\pi\,(\theta^f_L)_0^{\,0}
&\!\!=\!\!&
-\frac{f_L}{r^2}
-\frac{f_L'}{r}
\ ,
\\
\label{Tolmanf2}
8\pi\,(\theta^f_L)_1^{\,1}
&\!\!=\!\!&
-f_L\left(\frac{1}{r^2}+\frac{\xi'}{r}\right)
\ ,
\\
\label{Tolmanf3}
8\pi\,(\theta^f_L)_2^{\,2}
&\!\!=\!\!&
-\frac{f_L}{4}\left(2\xi''+\xi'^2+2\frac{\xi'}{r}\right)
- \frac{f'_L}{4}\left(\xi'+\frac{2}{r}\right)
\ .
\end{eqnarray}
It is clear from these equations that we must impose a constraint for $\theta_{\mu \nu}$ in order to find a solution for the $f_L$ and $\theta^f_L$ components. Then, the new solution is completely determined by the set of functions $\{\xi,\mu+\alpha_L f_L,T^0_{\mu \nu}+\alpha_L (\theta^f_L)_{\mu \nu}\}$. 

Now, in order to complete the left path, we shall use the gravitational decoupling method, restricted to a pure temporal metric deformation, taking as seed the solution obtained in the previous step. In this case, the system of equations that we need to solve is given by
\begin{eqnarray}
\label{Tolman1era2da1}
\hspace{-0.7cm}8\pi(\theta^h_L)_0^{0} & = & 0, \\
\label{Tolman1era2da2}
\hspace{-0.7cm} 8\pi(\theta^h_L)_1^{1} &=& - \frac{(\mu+\alpha_L f_L) h_L'}{r}, \\
\label{Tolman1era2da3}
\hspace{-0.7cm}
8\pi(\theta^h_L)_2^{2} &=& -\frac{\mu+\alpha_L f_L}{4} \Bigg(2h_L'' + \beta_L h_L'^2 + \frac{2h_L'}{r} + 2\xi'h_L'\Bigg) - \frac{(\mu'+\alpha_L f'_L)h_L'}{4}.
\end{eqnarray}
As before, in order to solve the system for $h_L$ and the $\theta^L_h$ components, we need to provide an additional constraint. The final solution is determined by the set of functions $\{\nu_L,\tilde{\mu}_L, T^L_{\mu\nu}\}$ where
\begin{eqnarray}
\nu_L & = & \xi+\beta_Lh_L, \\
\tilde{\mu}_L &=& \mu+\alpha_L f_L, \\
T^L_{\mu\nu} &=& \alpha_L (\theta^f_L)_{\mu \nu} + \beta_L (\theta^h_L)_{\mu \nu}.
\end{eqnarray}
Thus, given two constraints (for the two sources), we can find a solution with deformations in the spatial and temporal metric components.

\subsection{Right path}

On the right path, the first step is to use the gravitational decoupling method with pure temporal metric deformation, considering as seed the solution given by $\{\xi,\mu,T^0_{\mu \nu}\}$. Thus, the resulting system of equations that we need to solve is given by
\begin{eqnarray}
\label{Tolman1era2da11}
8\pi(\theta^h_R)_0^{0} & = & 0, \\
\label{Tolman1era2da22}
8\pi(\theta^h_R)_1^{1} &=& - \frac{\mu h_R'}{r}, \\
\label{Tolman1era2da33}
8\pi(\theta^h_R)_2^{2} &=& -\frac{\mu}{4} \left(2h_R'' + \alpha_R h_R'^2 + \frac{2h_R'}{r}+2\xi'h_R'\right) - \frac{\mu'h_R'}{4}.
\end{eqnarray}
As in the left path, we need to provide more information by hand in order to solve the system for $h_R$ and the $\theta^h_R$ components. Thus, the set of functions $\{\xi + \alpha_R,\mu,T^0_{\mu \nu}+\alpha_R (\theta^h_R)_{\mu \nu}\}$ fully characterizes the new solution of Einstein's equations. Now we can use this result as a seed solution for the gravitational decoupling method restricted to pure spatial deformation (MGD). In this case, it is a straightforward computation to show that the resulting system of equations is given by
\begin{eqnarray}
\label{Tolmanf11}
\hspace{-0.8cm} 8\pi\,(\theta^f_R)_0^{\,0}
&\!\!=\!\!&
-\frac{f_R}{r^2}
-\frac{f_R'}{r}
\ ,
\\
\label{Tolmanf22}
\hspace{-0.8cm} 8\pi\,(\theta^f_R)_1^{\,1}
&\!\!=\!\!&
-f_R\left(\frac{1}{r^2}+\frac{\xi'+\alpha_R h'_R}{r}\right)
\ ,
\\
\label{Tolmanf33}
\hspace{-0.8cm} 8\pi\,(\theta^f_L)_2^{\,2}
&\!\!=\!\!&
-\frac{f_L}{4}\Big(2(\xi''+\alpha_R h''_R)+(\xi'+\alpha_R h'_R)^2 + 2\frac{\xi'+\alpha_R h'_R}{r}\Big)
- \frac{f'_R}{4}\left(\xi'+\alpha_R h'_R+\frac{2}{r}\right)\hspace{-0.1cm}.
\end{eqnarray}
Here, it is clear that we also need to impose a constraint by hand in order to solve the system of equations. The set of functions $\{\nu_R,\tilde{\mu}_R, T^R_{\mu\nu}\}$ then completely determines the final solution of the right path
\begin{eqnarray}
\widetilde{\nu}_R & = & \xi+\alpha_Rh_R, \\
\tilde{\mu}_R &=& \mu+\beta_R f_R, \\
T^R_{\mu\nu} &=& \alpha_R (\theta^h_R)_{\mu \nu} + \beta_R (\theta^f_R)_{\mu \nu}.
\end{eqnarray}
Thus, given two constraints, we can resolve the two systems of equations to get a solution with both spatial and temporal metric deformations.

\subsection{Connection between the right and left path}

As we mentioned before, in order to solve the two systems of equations corresponding to the left and right paths, we need to provide further information by hand (as in EGD). This can be seen as the imposition of two constraints. It is clear that imposing different constraints on the left and right paths will lead to different solutions. However, if we impose the conditions
\begin{eqnarray}
\alpha_Rh_R = \beta_Lh_L, \quad \beta_R f_R = \alpha_L f_L, \label{constr_equiv}
\end{eqnarray}
it can be shown that
\begin{eqnarray}
\alpha_R(\theta_R^h)^0_0 &=& \beta_L (\theta_L^h)^0_0 = 0 , \\
\alpha_R (\theta_R^h)^1_1 &=& \beta_L (\theta_L^h)^1_1 + \frac{\beta_L\alpha_Lf_Lh'_L}{8\pi r}, \\
\alpha_R (\theta_R^h)^2_2 &=& \beta_L (\theta_L^h)^2_2 + \frac{\beta_L\alpha_Lf_L}{32\pi}\Big(2h_L'' +  \beta_L h_L'^2  +  \frac{2h_L'}{r}+2\xi'h_L' \Big) + \frac{\beta_L\alpha_Lf_L'h_L'}{32\pi},
\end{eqnarray}
and
\begin{eqnarray}
\beta_R (\theta^f_R)^0_0 &=& \alpha_L (\theta^f_L)^0_0, \\
 \beta_R (\theta^f_R)^1_1 & = & \alpha_L(\theta^f_L)^1_1 - \alpha_L\beta_L \frac{f_Lh'_L}{8\pi r}, \\
\beta_R (\theta^f_R)^2_2 & = &  \alpha_L(\theta^f_L)^2_2 -\frac{\alpha_L f_L}{32\pi} \Big(2h_L'' + \beta_L h_L'^2 +  \frac{2h_L'}{r}+2\xi'h_L'\Big) -\frac{(\alpha_L f'_L)h_L'}{32\pi} ,
\end{eqnarray}
will lead us to the same solution obtained with the left path, using the right path.

This gives us a relationship between the two paths. Indeed, these equations imply that every solution obtained using the left (right) path can also be obtained using the right (left) path. Thus, the paths are interchangeable, and we can therefore use the one that is most convenient in each case.

\section{Polytropic models}

In this section, we present the formalism to obtain analytic solutions of Einstein’s equations satisfying a polytropic equation of state from a known solution. For simplicity, we shall use the left path of the 2-step GD. Thus, the first step will be a deformation of the spatial metric component (MGD). In order to solve the system of equations given by (\ref{Tolmanf1})-(\ref{Tolmanf3}), we will impose the constraint
\begin{eqnarray}\label{Constraint1}
(\theta_L^f)^1_1 = \frac{1}{\alpha_L}P_r, \label{constraint1LP}
\end{eqnarray}
where $P_r$ is the radial pressure of the seed solution. We will obtain a solution with null radial pressure. Thus, from (\ref{Tolmanf2}) we obtain that the deformation function $f_L$ is given by
\begin{eqnarray}\label{sdefor}
f_L &=& - \frac{8\pi r^2 P_r}{\alpha_L(1+r\xi')}.
\end{eqnarray}
We can compute $(\theta^f_L)^0_0$, $(\theta^f_L)^2_2$ from (\ref{Tolmanf1}) and (\ref{Tolmanf3}), respectively, once we know $f_L$.

Following the 2-step GD procedure, we continue with the deformation of the temporal metric component, taking as seed the previous solution. In this case, we need to solve the system of equations (\ref{Tolman1era2da1})-(\ref{Tolman1era2da3}). Here we impose the constraint given by
\begin{eqnarray}\label{Constraint2}
(\theta_L^h)^1_1 = -\frac{K}{\beta_L}\widetilde{\rho}^{1+1/n} -\frac{G}{\beta_L}\widetilde{\rho} - \frac{D}{\beta_L}, \label{constraint2LP}
\end{eqnarray}
where $\widetilde{\rho} = \rho + \alpha_L (\theta^f_L)^0_0$, $\rho$ is the seed solution's energy density, $K$ is the polytropic constant, $n$ is the polytropic index and $G,D$ are constants. In this way, the final radial pressure will satisfy a polytropic equation of state
\begin{eqnarray}
\tilde{P}_r &=& P_r -\alpha_L (\theta_L^f)^1_1 - \beta_L (\theta_L^h)^1_1 
 =  K\widetilde{\rho}^{1+1/n} +G\widetilde{\rho} +D.
\end{eqnarray}
Therefore, it can be seen from (\ref{Tolman1era2da2}) that
\begin{eqnarray}\label{tdf}
h_L &=& \frac{8\pi}{\beta_L} \int dr \frac{r}{\mu+\alpha_L f_L}\left( K\widetilde{\rho}^{1+1/n} +G\widetilde{\rho} +D\right) ,
\end{eqnarray}
and from (\ref{Tolman1era2da3}) we can obtain $(\theta^h_L)^2_2$.

In order to show an example, let us consider the Tolman IV solution of Einstein equations as a seed solution
\begin{eqnarray}\label{Tolman}
e^\xi & = & B^2\left(1+\frac{r^2}{A^2}\right), \\
\mu & = & \frac{\left(1-\frac{r^2}{C^2}\right)\left(1+\frac{r^2}{A^2}\right)}{\left(1+\frac{2r^2}{A^2}\right)}, \\
\rho & = & \frac{3A^4+A^2(3C^2+7r^2)+2r^2(C^2+3r^2)}{8\pi C^2(A^2+2r^2)^2}, \\
P & = & \frac{C^2-A^2-3r^2}{8\pi C^2(A^2+2r^2)} ,
\end{eqnarray}
where $A$, $B$ and $C$ are constants.

Introducing these expressions in (\ref{sdefor}) we obtain
\begin{equation} \label{sdefor2}
f_L=\frac{(A^2+3r^2-C^2)r^2(A^2+r^2)}{\alpha C^2(A^2+2r^2)(A^2+3r^2)}.
\end{equation}
Thus, the solution of Einstein's equations with null radial pressure is characterized by the following functions
\begin{eqnarray}
e^\nu & = & e^\xi = B^2\left(1+\frac{r^2}{A^2}\right) , \label{g01}\\ 
\bar{\mu} &=&\mu +\alpha_L f_L=\frac{A^2+r^2}{A^2+3r^2}, \label{g02} \\
\bar{\rho} &=&\rho + \alpha_L (\theta^f_L)^0_0=\frac{3}{4\pi}\frac{(A^2+r^2)}{(A^2+3r^2)^2}, \label{g03} \\
\bar{P}_r & = &  P-\alpha_L (\theta^f_L)^1_1=0, \label{g04}\\
\bar{P}_t &=& P-\alpha_L (\theta^f_L)^2_2=\frac{3}{8\pi}\frac{r^2}{(A^2+3r^2)^2}. \label{g05}
\end{eqnarray} 
It is important to emphasize that, in this work, we are not interested in these kinds of solutions, since our main goal is the construction of polytropic matter distributions. However, it is worth mentioning that anisotropic solutions with null radial pressure may correspond to a distribution whose matter content consists of different shells, as, for example, the Einstein Cluster.

Finally, following the procedure presented at the beginning of this section, it is easy to find the following expression for $h_L$  
\begin{eqnarray}
\hspace{-0.5cm} h_L &=& \frac{8\pi}{\beta_L} \int \frac{r(A^2+3r^2)}{A^2+r^2}\Big[ K\bar{\rho}^{1+1/n}+ G\bar{\rho} +D\Big]  dr,
\end{eqnarray}
The final solution, which satisfies a polytropic equation of state, with deformations of the spatial and temporal metric components, is
\begin{eqnarray}\label{SolutionTol4}
e^{\bar{\nu}} &=&B^2\left(1+\frac{r^2}{A^2}\right)e^{\beta_L h_L}, \label{FinalSol1}\\
\widetilde{\mu} &=&  \mu + \alpha_L f_L , \label{FinalSol2} \\
\widetilde{\rho} &=& \rho + \alpha (\theta^f_L)^0_0 , \label{FinalSol3} \\
\widetilde{P}_r &=& P- \alpha_L (\theta^f_L)^1_1-\beta_L(\theta^h_L)^1_1 = K\widetilde{\rho}^{1+1/n} +G\widetilde{\rho}+ D, \label{FinalSol4} \\
\widetilde{P}_t &=&  P - \alpha_L(\theta^f_L)^2_2 - \beta_L(\theta^h_L)^2_2 .\label{FinalSol5}
\end{eqnarray}
Now, from the matching conditions (\ref{nursig})-(\ref{psig}) we obtain
\begin{eqnarray}
B^2 & = & \left(1-\frac{2M}{R}\right)\left(1+\frac{R^2}{A^2}\right)^{-1}e^{-\beta_L h_L(R)}\label{BMC}, \\
A^2 & = & \frac{R^2}{M}(R-3M), \label{AMC} \\
D & = & -K\left( \frac{3}{4\pi} \frac{A^2+R^2}{(A^2+3R^2)^2} \right)^{1+1/n} 
- \frac{3}{4\pi}G\frac{(A^2+R^2)}{(A^2+3R^2)^2}, \label{DMC}. 
\end{eqnarray}
respectively. Thus we can write 
\begin{eqnarray}
\widetilde{\rho} &=& \frac{3 M \left(M \left(r^2-3 R^2\right)+R^3\right)}{4 \pi  \left(3 M (r-R) (r+R)+R^3\right)^2}, \label{frho} \\
\widetilde{P}_r &=& K(\widetilde{\rho}^{1+1/n}-\widetilde{\rho}^{1+1/n}_\Sigma) +G(\widetilde{\rho}-\widetilde{\rho}_\Sigma)   \label{frpr}
\end{eqnarray}
and
\begin{eqnarray}
\widetilde{P}_t &=& \widetilde{P}_r + \frac{1}{8\pi n(R^3+3M(r^2-R^2))^3}\left[ 6GM^2nr^2(3M-R)R^2, \right. \nonumber \\
&+& \left. 3M^2nr^2(R^3+3M(r^2-R^2)) + 6GM^2nr^2(R^3+3M(r^2-R^2)) \right. \nonumber \\
&+& \left. 9G^2M^2nr^2(R^3+M(r^2-3R^2))+9k^2M^2n\widetilde{\rho}^{2/n}r^2(R^3+M(r^2-3R^2)) \right. \nonumber \\
&-& \left. 3GMn(6M(r^2-R^2))(R^3+M(r^2-3R^2)), \right. \nonumber \\
&-& \left.6KM\widetilde{\rho}^{1/n}r^2\left(4n\pi R^2(R-3M)^2\widetilde{\rho}_\Sigma(G+K\widetilde{\rho}_\Sigma^{1/n}) \right. \right. \nonumber \\
&+& \left. \left.  M(3M-R)R^2(2n+3Gn-5(1+n)-24n\pi r^2 \widetilde{\rho}_\Sigma(G+K\widetilde{\rho}_\Sigma^{1/n})) \right. \right. \nonumber \\
&+& \left. \left. 3M^2r^2(1-Gn+12n\pi r^2 \widetilde{\rho}_\Sigma(G+K\widetilde{\rho}_\Sigma^{1/n}) )\right) \right] \nonumber \\
&+& \frac{\widetilde{\rho}_\Sigma(G+K\widetilde{\rho}_\Sigma^{1/n})}{8\pi n(R^3+3M(r^2-R^2))(R^3+M(r^2-3R^2))}\left[ 24Mn\pi r^2(3M-R)R^2 \right. \nonumber \\
&-& \left. 4\pi n(3M^2r^4+R^4(R-3M)^2)- 8Mn\pi r^2(R^3+3M(r^2-R^2))\right. \nonumber \\
&-& \left. 24GMn\pi r^2 (R^3+M(r^2-3R^2)) + 4n\pi (R^3+3M(r^2-R^2))(R^3+M(r^2-3R^2)) \right. \nonumber \\
&+& \left. 16n\pi^2r^2(R^3+3M(r^2-R^2))^2 \widetilde{\rho}_\Sigma(G+K\widetilde{\rho}_\Sigma^{1/n})\right]
\end{eqnarray}

where $\widetilde{\rho}_\Sigma = \widetilde{\rho}(R)$. 

We omitted the expression for the tangential pressure because it is too long. However, it is important to notice that the final solution does not depend on $C$. Now, in order to analyze the behavior of our solution, it will be useful to define the following adimensional variables
\begin{eqnarray}
\hat{P}_r = \tilde{P}_r/P_0, \quad \hat{P}_t = \tilde{P}_t/P_0, \label{pdimless}
\end{eqnarray}
\begin{eqnarray}
\hat{\rho} = \tilde{\rho}/\rho_c, \quad \hat{m} = m/M, \quad q = P_0/\rho_c, \quad G = q \hat{G}, \label{rhodimless}
\end{eqnarray}
\begin{eqnarray}
x= r/R, \quad y = M/R, \quad D= P_0\hat{D}\label{xdimless}
\end{eqnarray}
with $P_0 = K\rho_c^{1+1/n}$.
It is important to mention that $q$ is a parameter that allows us to recover the Newtonian limit. Specifically, the limit $c\rightarrow \infty$ is equivalent to $q\rightarrow 0$.Then, using eq. (\ref{frho})-(\ref{frpr}), we can write
\begin{eqnarray}
\hat{\rho} & = & \frac{(1-3 y) \left(\left(x^2-3\right) y+1\right)}{\left(3 \left(x^2-1\right) y+1\right)^2}, \label{rhonor} \\
\hat{P}_r & = & \hat{\rho}^{1+1/n}-\hat{\rho}_\Sigma^{1+1/n} + \hat{G} (\hat{\rho}-\hat{\rho}_\Sigma), \label{pradnor}
\end{eqnarray}
and
\begin{eqnarray}
\hat{P}_t &=& \hat{P}_r + \frac{x}{2}\bigg[\left((n+1)\frac{\hat{\rho}^{1/n}}{n}+\hat{G}\right)\left(\frac{2 x y (3 y-1) \left(3 \left(x^2-5\right) y+5\right)}{\left(3 \left(x^2-1\right) y+1\right)^3}\right) \nonumber \\ &+& \frac{1}{q}\left(\frac{y}{1-3y}\right)\left(\frac{(1-3y)\hat{m}+3q\hat{P}_rx^3}{x^2(1-2\hat{m}y/x)}\right)(\hat{\rho}+q\hat{P}_r)\bigg],
\end{eqnarray}
where
\begin{eqnarray}
\hat{m} = \frac{x^3}{3 \left(x^2-1\right) y+1}. \label{adm}
\end{eqnarray}

Now, we can define an adimensional complexity factor given by
\begin{eqnarray}
\hat{Y}_{TF} = \frac{\lvert Y_{ TF} \rvert}{4\pi P_0} = \bigg\lvert 2\hat{\Delta} + \frac{1}{q}\left(\hat{\rho}-\frac{\hat{m}(1-3y)}{x^3}\right)\bigg\rvert. \label{Ytfnor}
\end{eqnarray}

\begin{figure}
    \centering
    \resizebox{0.45\textwidth}{!}{
    \includegraphics{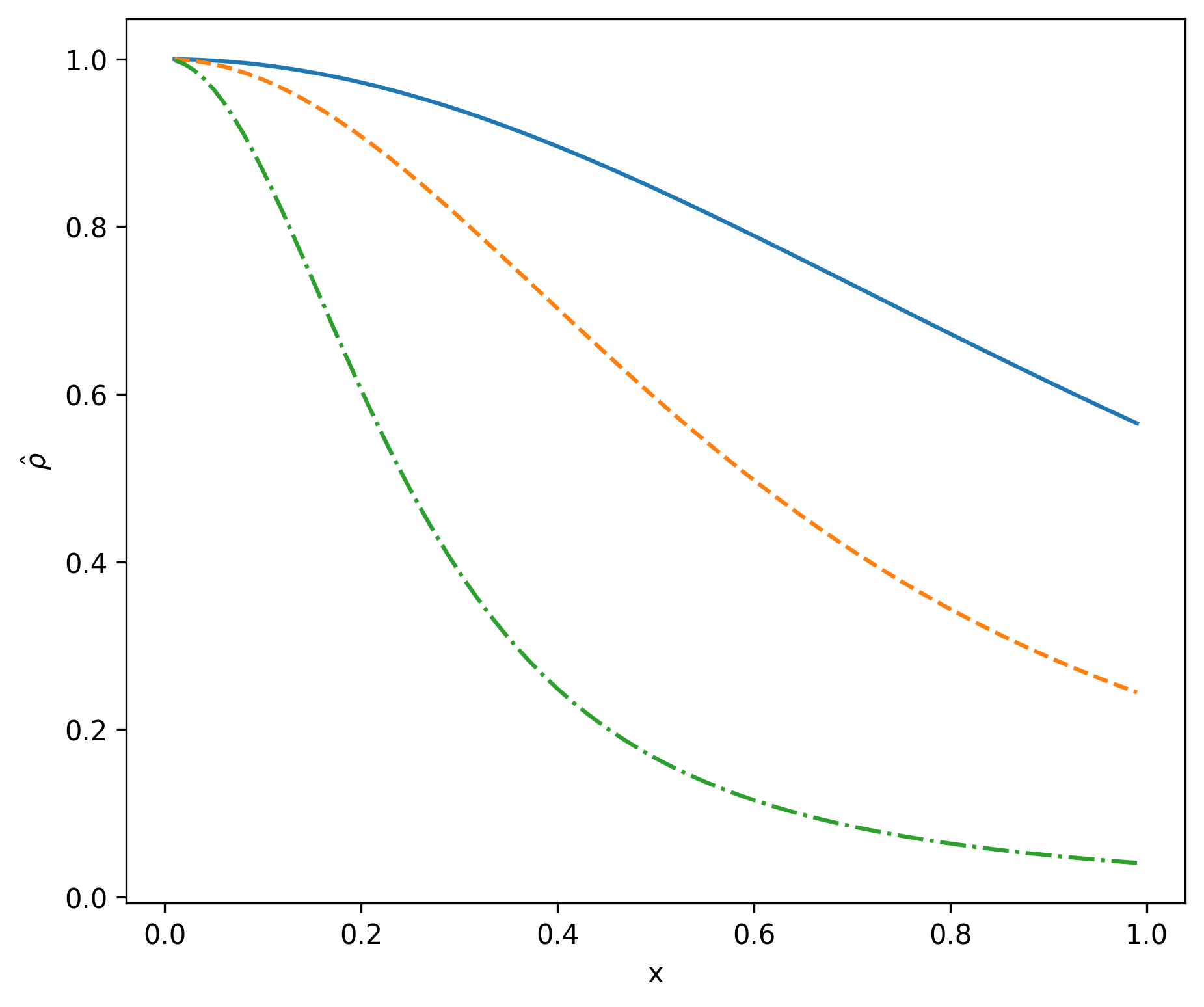}
    }
    \caption{$\hat{\rho}$ vs $x$ for $y=0.1$ blue (solid) curve, $y=0.2$ orange (dashed) curve and $y=0.3$ green (dot-dashed) line.}
    \label{fig:den}
\end{figure}

\begin{figure}
    \centering
    \resizebox{0.45\textwidth}{!}{
    \includegraphics{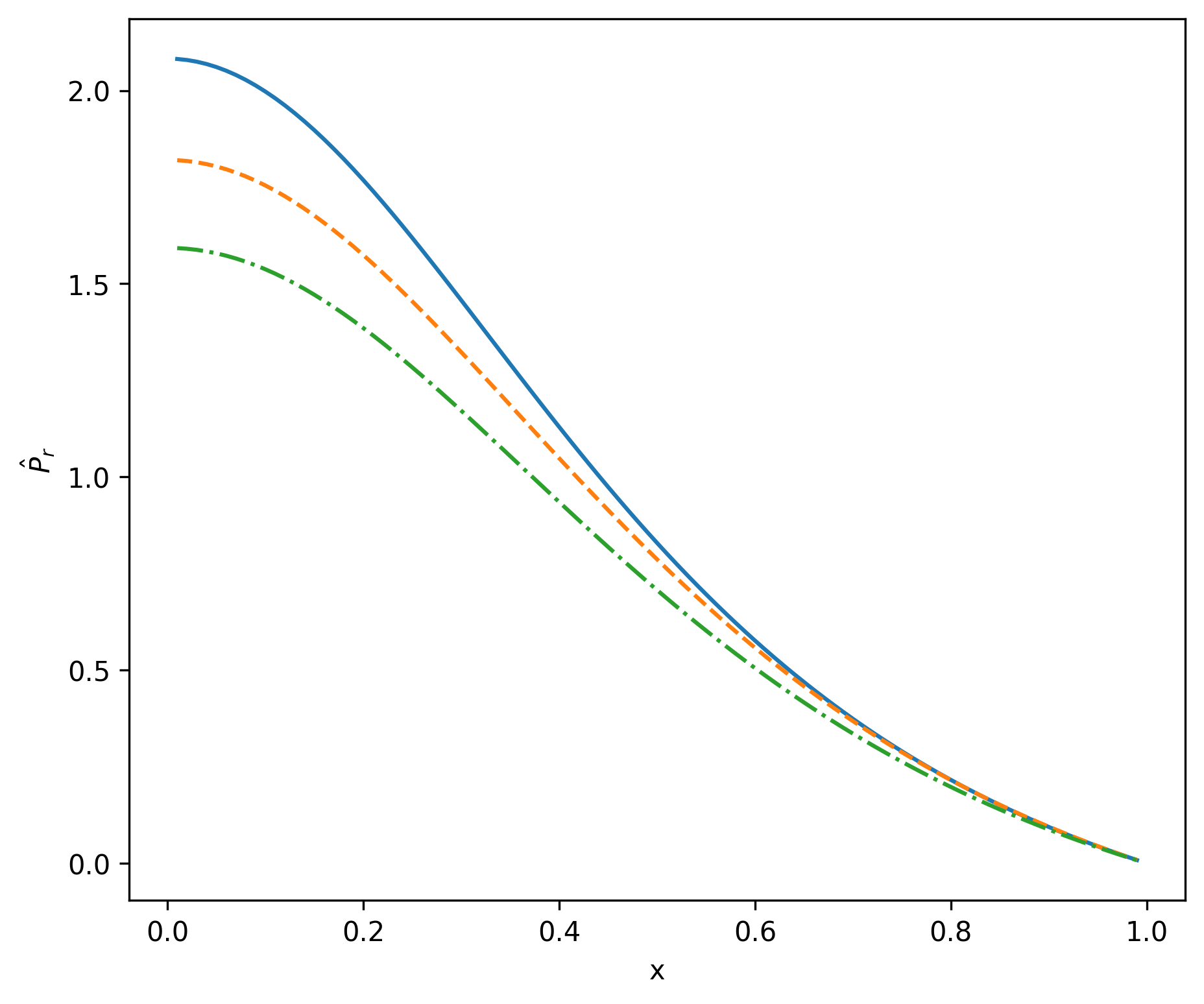}}
    \caption{$\hat{P}_r$ vs $x$ for $y=0.2$, $\hat{G}=1.5$, $q=0.03$ and $n=1$ blue (solid) curve, $n=2$ orange (dashed) curve and $n=3$ green (dot-dashed) line. The last two curves were consider with an scaling factor of $0.9$ and $0.8$, respectively.}
    \label{fig:prn}
\end{figure}

\begin{figure*}
    \centering
    \resizebox{0.8\textwidth}{!}{
    \includegraphics{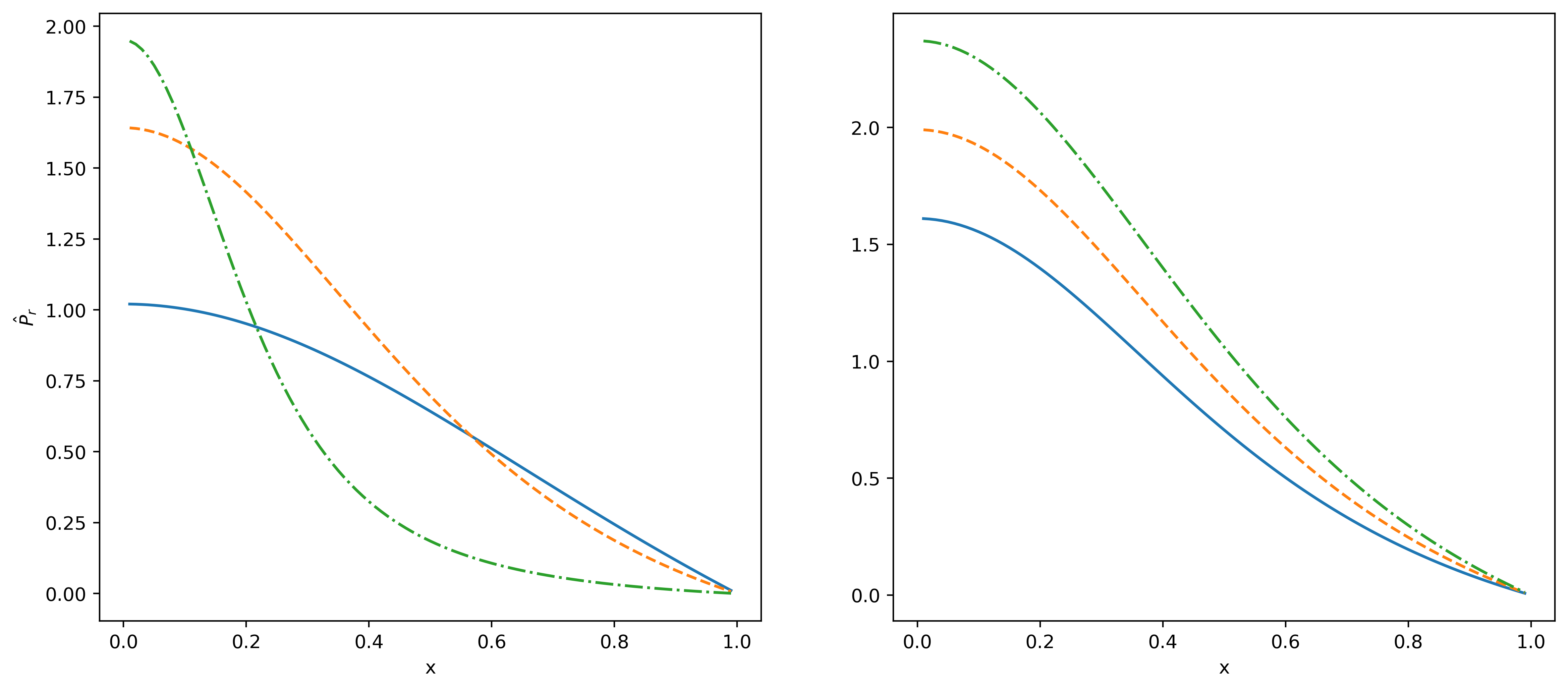}}
    \caption{$\hat{P}_r$ vs $x$ for (left panel) $\hat{G}=1.0$, $q=0.03$, $n=2.0$ for $y=0.1$ blue (solid) curve, $y=0.2$ orange (dashed) curve and $y=0.3$ green (dot-dashed) line. $\hat{P}_r$ vs $x$ (right panel) for $y=0.2$, $q=0.03$, $n=3.0$ and $\hat{G}=1.0$ blue (solid) curve, $\hat{G}=1.5$ orange (dashed) curve and $\hat{G}=2.0$ green (dot-dashed) line .}
    \label{fig:prgy}
\end{figure*}

\begin{figure*}
    \centering
    \resizebox{0.8\textwidth}{!}{
    \includegraphics{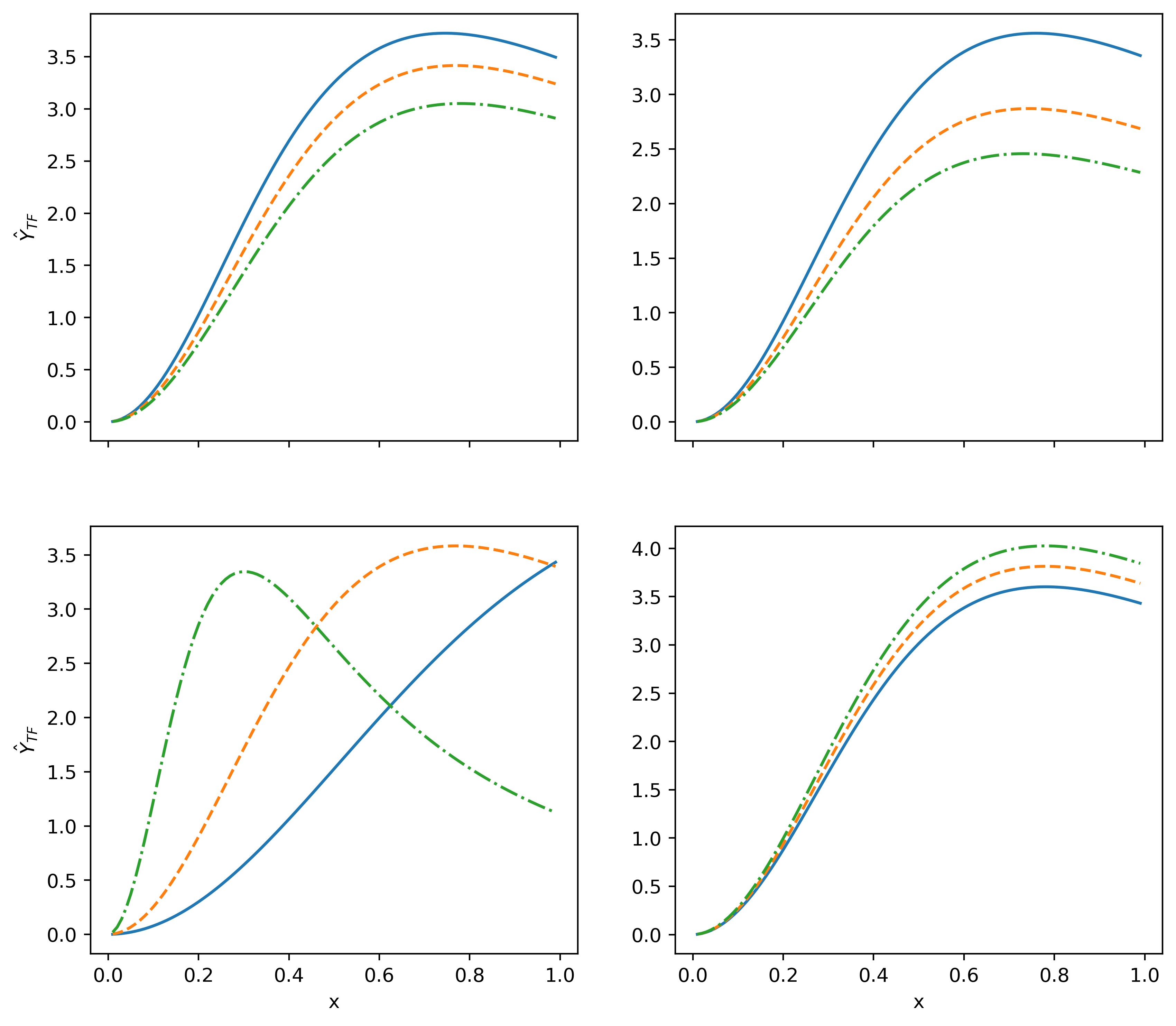}}
    \caption{$Y_{TF}$ vs $x$ for (top left) for $y=0.2$, $\hat{G}=1.5$, $q=0.03$ and $n=1$ blue (solid) curve, $n=2$ orange (dashed) curve and $n=3$ green (dot-dashed) line.  The last two curves were considered with an scaling factor of $0.9$ and $0.8$, respectively. $Y_{TF}$ vs $x$ (top right) for $\hat{G}=1.0$, $n=1.5$, $y=0.2$ and $q=0.03$ blue (solid) curve, $q=0.04$ orange (dashed) curve and $q=0.05$ green (dot-dashed) curve. $Y_{TF}$ vs $x$ for (bottom left) $\hat{G}=1.0$, $q=0.03$, $n=2.0$ for $y=0.1$ blue (solid) curve, $y=0.2$ orange (dashed) curve and $y=0.3$ green (dot-dashed) line. $Y_{TF}$ vs $x$ for (bottom right) for $y=0.2$, $q=0.03$, $n=3.0$ and $\hat{G}=1.0$ blue (solid) curve, $\hat{G}=1.5$ orange (dashed) curve and $\hat{G}=2.0$ green (dot-dashed) line. }
    \label{fig:complexity}
\end{figure*}

To examine the properties of this solution in greater depth, we present the behavior of the different physical variables for various values of the parameters $y$, $q$, $n$ and  $\hat{G}$ in the figures (\ref{fig:den})-(\ref{fig:complexity}). We verify that, for all the chosen values, the solution meets the physical acceptability conditions presented in Appendix A. 
Although the general analysis of the physical acceptability conditions is very complicated because of the high complexity of some physical variable expressions, we can analyze some of the aforementioned conditions.

It is clear from (\ref{rhonor}) that $y\neq 1/3$ is required to avoid a singularity at the center of the distribution. Notice that this condition also ensures that $\hat{P}_r$ and $\hat{P}_t$ are free of singularities at the center of the distribution. Furthermore, it is easy to prove that
\begin{eqnarray}
\begin{array}{c}
   \widetilde{\rho} \geq 0    \\
    \partial_r \widetilde{\rho} \leq 0 
\end{array} \Longrightarrow y < 1/3.
\end{eqnarray}
On the other hand, using (\ref{rhonor}) it is easy to find that 
\begin{eqnarray}
\frac{\partial M}{\partial \widetilde{\rho}_c} = \frac{4\pi R^3}{3(1+4\pi \widetilde{\rho}_c R^2)^2} \geq 0,
\end{eqnarray}
which implies that our model satisfies the Harrison-Zeldovich-Novikov stability condition. 
\begin{figure*}
    \centering
    \resizebox{0.8\textwidth}{!}{
    \includegraphics{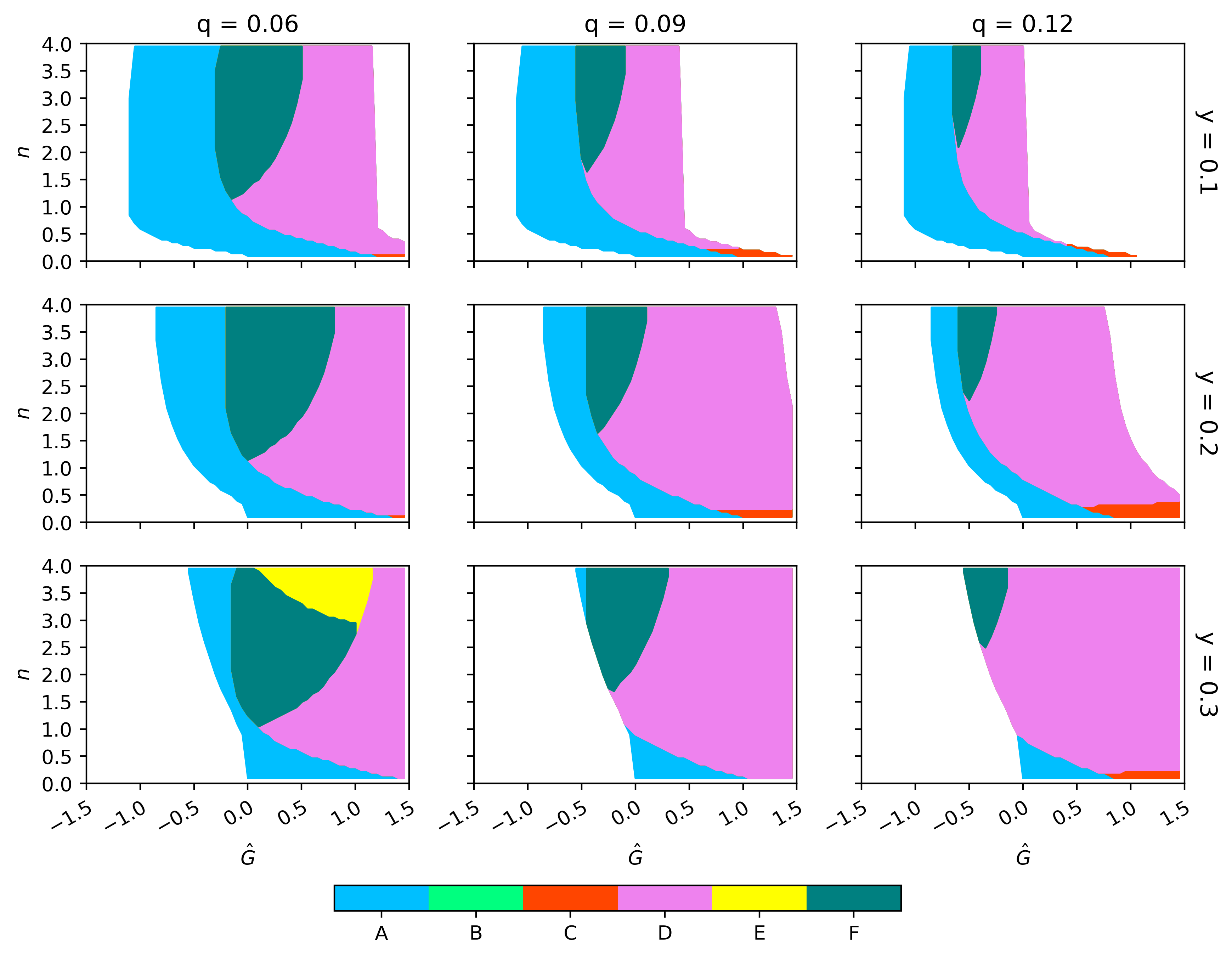}}
    \caption{Tuples ($n,\hat{G}$) that satisfy the acceptability conditions in appendix A. Each color in this graphic represent a set of acceptability's conditions that are satisfied. This are: blue region (A) only condition 1, green region (B) conditions 1-2, red region (C) conditions 1-3, violet region (D) conditions 1-4, yellow region (E) conditions 1-5 and dark green region (F) conditions 1-6.}
    \label{fig:nG1}
\end{figure*}
\begin{figure*}
    \centering
    \resizebox{0.8\textwidth}{!}{
    \includegraphics{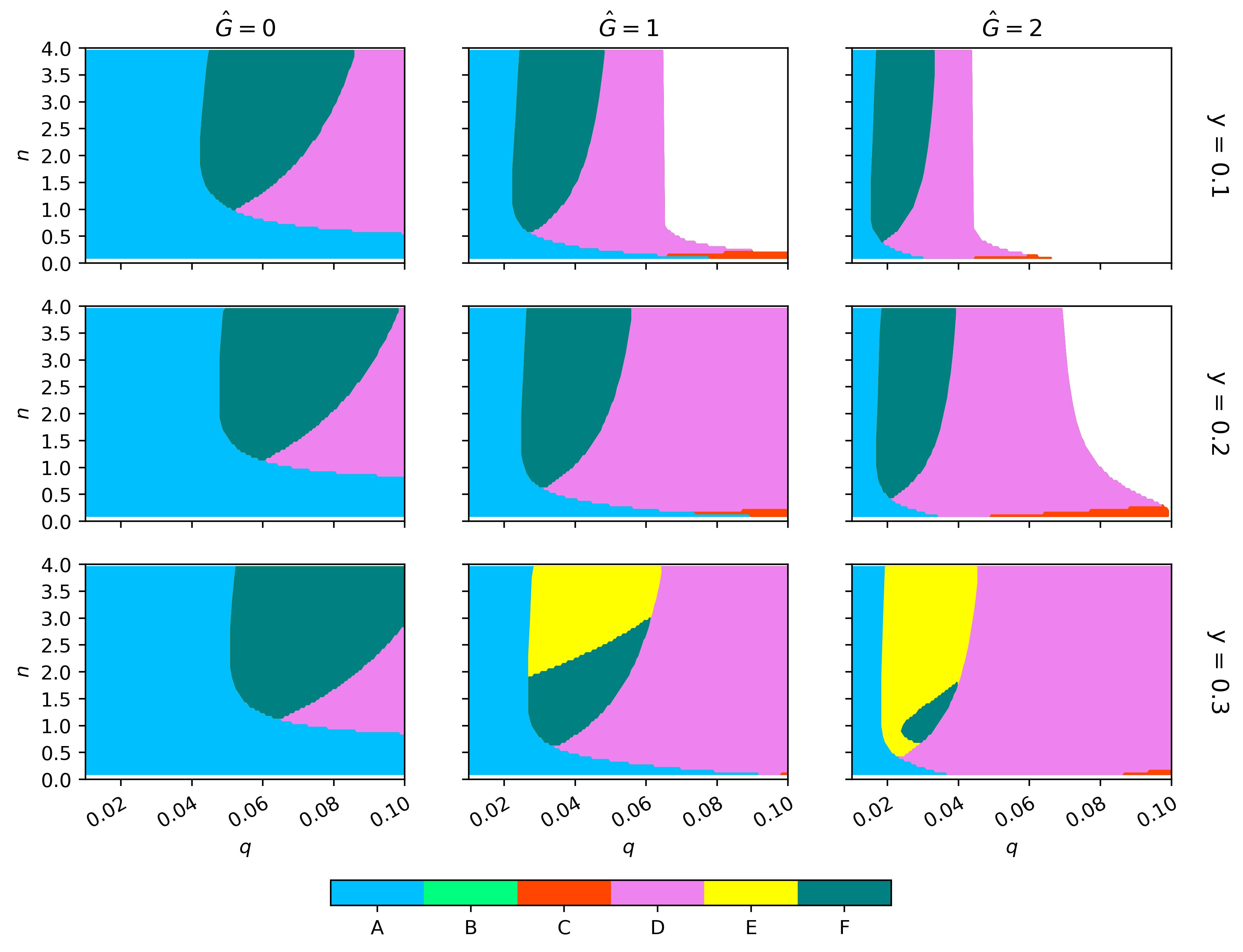}}
    \caption{Tuples ($n,q$) that satisfy the acceptability conditions in appendix A. Each color in this graphic represent a set of acceptability's conditions that are satisfied. This are: blue region (A) only condition 1, green region (B) conditions 1-2, red region (C) conditions 1-3, violet region (D) conditions 1-4, yellow region (E) conditions 1-5 and dark green region (F) conditions 1-6.}
    \label{fig:nq2}
\end{figure*}

Before continuing with the analysis of the results, we shall make some comments. First, the choice of the left path to get polytropic solutions to Einstein’s equations is only for simplicity. Indeed, from the results of section 4.3, it is always possible to find the same solution using the right path. On the other hand, one can ask if it is possible to obtain polytropic solutions with a sequence of pure spatial or temporal deformations of the metric. In the case of pure spatial metric deformations, imposing a polytropic equation of state in the final solution implies a highly non-linear differential equation for $f$, which does not have analytic solutions (in general). Thus, for these models, the MGD method does not pose any important advantage over a pure numeric analysis. In the case of pure temporal deformations, it is possible to get analytical polytropic solutions. However, since these kinds of metric deformations do not change the energy density, this procedure corresponds to solving the TOV equation for $\Delta$ after introducing the energy density expression of the seed solution.

\section{Results}
The solution given by (\ref{FinalSol1})-(\ref{FinalSol5}) and (\ref{BMC})-(\ref{DMC}), corresponds to an anisotropic polytropic solution of Einstein equations obtained by 2-step GD using Tolman IV as seed solution. In figures (\ref{fig:den})-(\ref{fig:complexity}) we present the behaviour of the dimensionless physical variables $\hat{\rho}$, $\hat{P}_r$ and $\hat{Y}_{TF}$ given by (\ref{rhonor}), (\ref{pradnor}) and (\ref{Ytfnor}), respectively, for different values of $y$, $n$, $q$ and $\hat{G}$ given by (\ref{rhodimless})-(\ref{xdimless}). The polytropic constant $K$ and $G$ from expression (\ref{polytropicEoS}) are encoded in $q$ and $\hat{G}$, respectively. For all the cases considered, we have chosen $y<1/3$ such that it ensures the good behaviour of the energy density of the distribution. For fixed $M$, smaller values of $y$ will correspond to bigger distributions. We have verified that in all the cases under the scope of this study, the physical acceptability conditions from appendix A are satisfied. Moreover, we have obtained a family of values for the free parameters where we can ensure the validity of the physical acceptability conditions for $0\leq r \leq R$ (or equivalently, $0\leq x \leq 1$). This is described by figures (\ref{fig:nG1}) and (\ref{fig:nq2}) and it will be discussed at the end of this section.

By construction, the final density obtained by the method proposed in this work is independent of $n$, $\hat{G}$ or $q$. Thus, for the chosen seed solution, $\hat{\rho}$ only depends on the values of $y$ and $x$ (see eq. (\ref{rhonor})). These dimensionless variables (\ref{pdimless})-(\ref{xdimless}) are usually used in the analysis of polytropes or contraction of fluid spheres in GR. From figure (\ref{fig:den}) we see that it behaves as a monotonically decreasing function of $x$ for the chosen values of $y\in[0.1,0.3]$. For bigger values of $y$, the normalized density decreases faster, leading to smaller values at the surface of the distribution. As it is expected, for fixed values of $M$ the density of energy in the center of the distribution ($\rho_c$) is smaller for bigger values of $R$. 

The good behaviour of the radial pressure is shown in figures (\ref{fig:prn}) and (\ref{fig:prgy}). From the former, the normalized radial pressure is plotted for $y=0.2$, $\hat{G}=1.5$, $q=0.03$ and the chosen values of $n\in[1.0,3.0]$. Although a scaling factor was considered for $n=2$ and $n=3$, we can see that the value of $\hat{P}_r$ at the center of the distribution grows with smaller values of $n$. From the left panel of figure (\ref{fig:prgy}), we can see the behavior of the dimensionless pressure for $\hat{G}=1.0$, $q=0.03$, $n=2.0$ and the chosen values of $y\in[0.1,0.3]$. While the value of $\hat{P}_r$ at the center grows with $y$, the bigger the value of $y$ it decreases faster towards the zero value at the surface. On the right panel we have that $\hat{P}_r$ is plotted for $y=0.2$, $q=0.03$, $n=3.0$ and $\hat{G}\in[1.0,2.0]$. The normalized radial pressure at the center of the distribution grows with $\hat{G}$. The anisotropy function is zero at the center and it increases near the surface of the distribution for all the cases considered, and the matching conditions ensure that the radial pressure vanishes at the surface of the distribution.

From figure (\ref{fig:complexity}) we can see the behaviour of the normalized complexity factor for different values of the free parameters. On the top left panel, $\hat{Y}_{TF}$ is plotted for $y=0.2$, $\hat{G}=1.5$, $q=0.03$ and selected values of $n\in[1.0,3.0]$. A scaling factor was considered for the last two curves corresponding to $n=2.0$ and $n=3.0$. We can see that the complexity factor grows with smaller values $n$ and it reach it maximum inside the distribution. A similar behaviour can be seen from the top right panel, where $\hat{Y}_{TF}$ is plotted for $\hat{G}=1.0$, $n=1.5$, $y=0.2$ and the chosen values of $q\in [0.03,0.05]$. The normalized complexity factor is bigger for smaller values of $q$ and it reach it maximum inside the distribution. On the bottom left panel, $\hat{Y}_{TF}$ is plotted for $\hat{G}=1.0$, $q=0.03$, $n=2.0$ and selected values of $y\in [0.1,0.3]$. While $\hat{Y}_{TF}$ increase towards the surface for $y=0.1$, for bigger values of $y$ the normalized complexity factor reach it maximum closer to the center of the distribution. The bigger the value of $y\leq 0.3$ the smaller the maximum of $\hat{Y}_{TF}$ and also it value at the surface of the distribution. Finally, in the bottom right panel, the normalized complexity factor is plotted for $y=0.2$, $q=0.03$, $n=3.0$ and $\hat{G}\in[1.0,2.0]$. We have that the complexity factor grows with $\hat{G}$ and it reach it maximum inside the distribution.

We have verified that the acceptability conditions from appendix A are satisfied for all the cases previously considered. Moreover, from figures (\ref{fig:nG1}) and (\ref{fig:nq2}) we have obtained a family of values for the tuples ($n,\hat{G}$) and ($n,q$) respectively, such that they satisfy the physical acceptability conditions from appendix A. We can see from figure (\ref{fig:nG1}) that $q=0.06$, $q=0.09$ and $q=0.12$ from left to right and $y=0.1$, $y=0.2$ and $y=0.3$ from top to bottom. Each point represents a pair of values $(n,\hat{G})$ and each color represents a set of acceptability conditions that are being satisfied for $0\leq x \leq 1$. The absence of the green region (B) in the figure shows us that the monotonically decreasing behaviour of the thermodynamical variables is guaranteed by conditions 3,4,5 and 6. The red region (C) corresponds to the values of ($n,\hat{G}$) where the dominant energy condition is not ensured by conditions 4,5 and 6. For $y=0.2$ and $y=0.3$ it can be seen it tends to increase its area for bigger values of $q$. The yellow region (E) is associated with a model that satisfies conditions 1 to 5 from appendix A, but violates the adiabatic index stability criteria. The dark green region corresponds to values of $(n,\hat{G})$ that satisfy the complete set of conditions of appendix A. It tends to decrease for bigger values of $q$ and $y$. It can be seen that this region is shifted to the left for bigger values of $q$ and for all $y$ under consideration. The case $y=0.2$ contains the situation where $M=1$ and $R=5$, which are typical values of high-density objects such as neutron stars, while $y=0.3$ is closer to the critical value $y=1/3$. For fixed $q$, we see that the amount of tuples $(n,\hat{G})$ that will satisfy all the physical conditions tends to increase for bigger values of $y$. Let us notice that, for $n<1.0$, we hardly find tuples related to solutions where the local anisotropy is positive. Moreover, for $\hat{G}<-0.5$, we have that the complete set of acceptability conditions is satisfied for bigger values of $q$ and smaller of $y$. In general, if we relax the condition on the local anisotropy, the number of values related to physical solutions increases. Finally, the case where the dominant energy condition is violated, is only characterized by the positivity and finiteness of the thermodynamical variables within the distribution (weak energy condition). Therefore, those exotic sources are not described by this model for the values of $n$ and $\hat{G}$ considered.

From figure (\ref{fig:nq2}), we can see that $\hat{G}=0$, $\hat{G}=1$ and $\hat{G}=2$ from left to right and $y=0.1$, $y=0.2$ and $y=0.3$ from top to bottom. Each point represents a pair of values $(n,q)$ where the acceptability conditions are fulfilled. As before, we notice that, because of the absence of the green color (B), the monotonically decreasing behavior of the pressures and the energy density is guaranteed by conditions 3,4,5 and 6. The dominant energy condition is ensured by the conditions 4,5 and 6, except for the small red region (C). For $y=0.2$ and $y=0.3$ it can be seen it tends to increase it area for bigger values of $G$. The condition on the local anisotropy is ensured by the adiabatic index stability criteria, except by the yellow region (E). On the other hand, the dark green region (F), where the complete set of acceptability conditions is satisfied, seems to be smaller for fixed $y$ and bigger $\hat{G}$. Here, the interval of values of $n$ tends to increase for fixed $y$ and bigger $\hat{G}$, except for $y=0.3$ where the interval is shifted to $n<2$. Moreover, the interval of values of $q$ tends to decrease for bigger values of $\hat{G}$ and is shifted to the left for all $y$ under consideration. We also notice that for $y=0.3$ and all values of $\hat{G}$ (or equivalently, for $\hat{G}=0$ and all values of $y$) all the tuples ($n,q$) under considerations will be related with, at least, one of the physical acceptability conditions. In this case, we can hardly find some tuples related to $n<0.5$ that will lead to physical solutions. As before, relaxing the condition on the local anisotropy will increase the number of tuples.

Summarizing our results, we found that with Tolman IV as seed solution: 
\begin{itemize}
\item Smaller values of $y$ leads to smaller values of $\hat{P}_r$ and to bigger values of $\hat{Y}_{TF}$ at the surface of the distribution.
 \item An increment of $n$ implies decrements of $\hat{P}_r$ and $\hat{Y}_{TF}$, increments of $\hat{G}$ implies increments of $\hat{P}_r$ and $\hat{Y}_{TF}$.
 \item For fixed $y$, the amount of tuples ($n,\hat{G}$) and ($n,q$) where the complete set of acceptability conditions are satisfied, tends to decrease for bigger $q$ and $\hat{G}$, respectively.
\end{itemize}

\section{Other models}
In this section, we shall present other polytropic models that could be constructed with the formalism presented in this work. Throughout this manuscript, we implemented the master polytropic state equation with the energy density. It is known, however, that we can also use a version of the polytropic state equation with the barytropic mass density $\rho_0$, which is given by 
\begin{eqnarray} \label{p2}
P_r = K \rho_0^{1+1/n} + G \rho_0 + D.
\end{eqnarray}
Because the baryonic mass density does not appear in Einstein's equations, we must provide the relationship between$\rho_0$ and $\rho$. The relation can be obtained by using the first and second laws of thermodynamics, which, for an adiabatic process, can be written as
\begin{eqnarray} \label{fslt}
d\left(\frac{\rho}{\mathcal{N}}\right)+ P_r d\left(\frac{1}{\mathcal{N}}\right) =0,
\end{eqnarray}
where $\mathcal{N}$ is the number of particle density, given by
\begin{eqnarray}
\mathcal{N} = \rho_0/m_0.
\end{eqnarray}
Thus, introducing (\ref{p2}) in (\ref{fslt}) we get the following 
\begin{eqnarray}
d\left(\frac{\rho}{\rho_0}\right) = (K\rho_0^{1/n-1}+G\rho_0^{-1}+D\rho_0^{-2})d\rho_0,
\end{eqnarray}
whose solution is given by 
\begin{eqnarray}
\rho = nK \rho_0^{1+1/n}+G\rho_0 \ln(\rho_0)-D +C\rho_0,
\end{eqnarray}
for $1/n \not= 0$ and
\begin{eqnarray}
\rho = (K+G)\rho_0 \ln(\rho_0)-D +C\rho_0,
\end{eqnarray}
for $1/n = 0$. In both cases, $C$ is an integration constant. In the Newtonian limit $\rho=\rho_0$, then we should set $C=1$ in all the cases. Thus, the final expressions are
\begin{eqnarray}
\rho &=& nK \rho_0^{1+1/n}+G\rho_0 \ln(\rho_0)-D +\rho_0, \quad \mbox{for} \quad 1/n \not= 0, \label{rl1} \\
\rho &=& (K+G)\rho_0 \ln(\rho_0)-D +\rho_0,  \quad \mbox{for} \quad 1/n = 0. \label{rl2}
\end{eqnarray}
Now, in order to apply the procedure presented in this work, with the state equation (\ref{p2}), we will only need to replace $\rho$ by $\rho_0$ in eq. (\ref{Constraint2}). Then, by using the relation (\ref{rl1}) or (\ref{rl2}), we could write the barionyc mass density in terms of the energy density and, in this way, close the system of equations. It is important to mention that the main goal of this section is not to write the full set of equations for this case. Instead, we want only to show how the models satisfying (\ref{p2}) could be constructed with the procedure presented in this manuscript. Finally, we would like to mention that, because of (\ref{rl1}) or (\ref{rl2}), it is probable that only a few values of $n$ will lead to an analytical solution of Einstein’s equations. At least with the procedure presented in this work.

\section{Conclusions}
The extended geometric deformation allows us to generate anisotropic solutions to the Einstein equations. It is characterized by considering deformation of both the temporal and radial components of the metric. However, we know that the system of equations of the EGD approach may be hard to solve. In the particular case where the deformations are not simultaneous, but consecutive, we have that there is a simpler way to obtain solutions of the Einstein equations. This method was named 2-step GD, and it provides two equivalent ways to generate new anisotropic solutions. They were named the left and right path. Although solutions obtained by the left path can also be obtained from the right path by imposing (\ref{constr_equiv}), one way may be more convenient than the other in particular cases. The consecutive deformations of the line element, are based on the assumption that the gravitational source $\theta_{\mu \nu}$ can be decomposed into two parts. One part will be responsible for the spatial metric deformation and the other one for the temporal metric deformation. Now, it is important to mention that the source related to the spatial deformation could have a physical interpretation, but the one related to the temporal deformation does not. Therefore, at least from the point of view of the 2-step GD, when both metric deformations are taken into account, it is better to consider that only the final source will have a physical meaning. Thus, only the final solutions with both deformations will be physically relevant.


The 2-step GD allows us to obtain analytic anisotropic solutions of Einstein equations satisfying a polytropic state equation. We choose for simplicity the left path of the 2-step GD, which considers first the radial deformation of the metric. In fact, imposition of the constraint given by (\ref{constraint1LP}) leads us to a solution with null radial pressure. Although there are certain situations where this may be of interest, we were only interested in the final solution. After the radial deformation, we may impose a polytropic constraint given by (\ref{constraint2LP}). In order to give an example, we consider Tolman IV as a seed solution. The resulting solution given by (\ref{FinalSol1})-(\ref{FinalSol5}), satisfies a generalized polytropic equation of state as it can be checked from (\ref{FinalSol4}). We analyze the behaviour of the thermodinamical variables and the complexity factor for different values of $y$, $x$, $q$, $\hat{G}$ and $n$. In addition, we check that our model satisfies the physical acceptability  conditions given in appendix A for a wide range of the parameters. In the final section, we also present how other polytropic models, in which the state equation is written in terms of the baryonic mass density, could be analyzed following the proposal of this work.

Finally, although we were not interested in model any particular scenario, we want to end this work by mentioning the possible applications of our results. As we mentioned in the introduction, the polytropic equation of state has a wide range of applications to studying the internal structure of self-gravitating objects in many circumstances. Indeed, solutions of Einstein’s equations can be very useful for modeling white dwarfs and analyzing the Chandrasekar mass limit. On the other hand, polytropic models may be relevant in the Schonberg-Chandrasekar limit. Moreover, the inclusion of local anisotropy in the pressures allows us to analyse more realistic objects. Now, it is important to mention that the intense magnetic fields observed in neutron stars and white dwarfs could break the spherical symmetry of the object. Thus, in this case, our results can be taken only as an approximation.

\backmatter

\bmhead{Acknowledgments}

P.L and C.L.H wants to say thanks for the financial support received by the Projects MINEDUC-UA ANT1956 and MINEDUC-UA ANT2156 of the Universidad de Antofagasta. P. L is grateful to CONICYT PFCHA / DOCTORADO BECAS CHILE/2019 - 21190517. C.L.H was supported by CONICYT PFCHA/DOCTORADO BECAS CHILE/2019-21190263. P.L and C.L.H also thanks to Semillero de Investigaci\'on SEM 18-02 from Universidad de Antofagasta and to the international ICTP Network NT08 for kind support. 

\bmhead{Data availibility statement} We have not used any data for the analysis presented in this work.

\begin{appendices}

\section{Physical acceptability conditions}\label{sec:AA}

In order to ensure that the solutions of Einstein’s equations describe realistic matter distributions, they should satisfy a series of physical conditions. Theses conditions are 
\begin{enumerate}
\item $P_r$, $P_t$ and $\rho$ are positive and finite inside the distribution.
\item $\frac{dP_r}{dr}$, $\frac{dP_t}{dr}$ and $\frac{d\rho}{dr}$ are monotonically decreasing.
\item Dominant energy condition: $0 \leq \frac{P_r}{\rho}\leq 1$ and $0 \leq \frac{P_t}{\rho}$ $\leq 1$.
\item Causality condition: $0<\frac{dP_r}{d\rho}<1$\hspace{0.1cm}, \hspace{0.1cm}$0<\frac{dP_t}{d\rho}<1$.
\item  The local anisotropy of the distribution should be zero at the center and positive towards the surface.
\item Adiabatic index stability criterion: $\frac{\rho + P_r}{P_r} \left(\frac{dP_r}{d\rho}\right)\geq \frac{4}{3}$. 
\item Harrison-Zeldovich-Novikov stability condition:
$\frac{dM}{d\rho_c} \geq 0$. 
\end{enumerate}
It can be seen that condition 1 ensures that the weak energy condition
\begin{eqnarray}
\rho + P_r \geq 0 \quad , \quad \rho + P_t \geq 0
\end{eqnarray}
is satisfied.




\end{appendices}



\begin{thebibliography}{100}

\bibitem{Chandrasekhar}
S.~Chandrasekhar.
\newblock {\em An Introduction to the Study of Stellar Structure}.
\newblock Astrophysical monographs. University of Chicago Press, 1939.

\bibitem{Schwarzschild}
M.~Schwarzschild.
\newblock {\em Structure and Evolution of Stars}.
\newblock Princeton University Press, 2015.

\bibitem{shapiro2}
{S. A.} Teukolsky and {S. L.} Shapiro.
\newblock {\em Black holes, white dwarfs, and neutron stars : the physics of
  compact objects}.
\newblock Wiley, 1983.

\bibitem{kippenhahn}
R.~Kippenhahn, A.~Weigert, and A.~Weiss.
\newblock {\em Stellar Structure and Evolution}.
\newblock Astronomy and Astrophysics Library. Springer Berlin Heidelberg, 2012.

\bibitem{Kovetz}
A.~Kovetz.
\newblock Slowly rotating polytropes.
\newblock {\em Astrophys. J.}, 154, 1968.

\bibitem{Goldreich}
P.~{Goldreich} and S.~V. {Weber}.
\newblock {Homologously collapsing stellar cores}.
\newblock {\em apj}, 238:991--997, June 1980.

\bibitem{Abramowicz}
M.~A. {Abramowicz}.
\newblock {Polytropes in N-dimensional spaces}.
\newblock {\em actaa}, 33(2):313--318, January 1983.

\bibitem{Tooper}
R.~F. {Tooper}.
\newblock {General Relativistic Polytropic Fluid Spheres.}
\newblock {\em apj}, 140:434, August 1964.

\bibitem{Bludman}
S.~A. {Bludman}.
\newblock {Stability of General-Relativistic Polytropes}.
\newblock {\em apj}, 183:637--648, July 1973.

\bibitem{Nilsson}
U.~S. {Nilsson} and C.~{Uggla}.
\newblock {General Relativistic Stars: Polytropic Equations of State}.
\newblock {\em Annals of Physics}, 286(2):292--319, December 2000.

\bibitem{Herrera9}
L.~Herrera and W.~O. Barreto.
\newblock Evolution of relativistic polytropes in the post–quasi–static
  regime.
\newblock {\em General Relativity and Gravitation}, 36:127--150, 2004.

\bibitem{Lai}
X.~Y. Lai and R.~X. Xu.
\newblock {A Polytropic Model of Quark Stars}.
\newblock {\em Astropart. Phys.}, 31:128--134, 2009.

\bibitem{Thirukkanesh}
S.~{Thirukkanesh} and F.~C. {Ragel}.
\newblock {Exact anisotropic sphere with polytropic equation of state}.
\newblock {\em Pramana}, 78(5):687--696, May 2012.

\bibitem{Shojai}
F.~{Shojai}, M.~R. {Fazel}, A.~{Stepanian}, and M.~{Kohandel}.
\newblock {On the Newtonian anisotropic configurations}.
\newblock {\em European Physical Journal C}, 75:250, June 2015.

\bibitem{zdenek1}
Z.~Stuchlek, S.~Hled\'ik, and J.~Novotn\'y.
\newblock {General relativistic polytropes with a repulsive cosmological
  constant}.
\newblock {\em Phys. Rev.}, D94(10):103513, 2016.

\bibitem{Herrera12}
L.~Herrera and W.~Barreto.
\newblock {Newtonian polytropes for anisotropic matter: General framework and
  applications}.
\newblock {\em Phys. Rev. D}, 87(8):087303, 2013.

\bibitem{Herrera13}
L.~Herrera and W.~Barreto.
\newblock {General relativistic polytropes for anisotropic matter: The general
  formalism and applications}.
\newblock {\em Phys. Rev. D}, 88(8):084022, 2013.

\bibitem{Herrera14}
L.~Herrera, A.~Di~Prisco, W.~Barreto, and J.~Ospino.
\newblock {Conformally flat polytropes for anisotropic matter}.
\newblock {\em Gen. Rel. Grav.}, 46(12):1827, 2014.

\bibitem{Chavanis}
P.~H.~Chavanis,
Astron. Astrophys. \textbf{537}, A127 (2012)
doi:10.1051/0004-6361/201116905
[arXiv:1103.2698 [astro-ph.CO]].

\bibitem{Abellan2}
G.~Abellán, E.~Fuenmayor, and L.~Herrera.
\newblock The double polytrope for anisotropic matter: Newtonian case.
\newblock {\em Physics of the Dark Universe}, 28:100549, 2020.

\bibitem{Abellan3}
G.~Abellan, E.~Fuenmayor, E.~Contreras, and L.~Herrera.
\newblock {The general relativistic double polytrope for anisotropic matter}.
\newblock {\em Phys. Dark Univ.}, 30:100632, 2020.

\bibitem{Contreras17}
A.~Ramos, C.~Arias, E.~Fuenmayor, and E.~Contreras.
\newblock {Class I polytropes for anisotropic matter}.
\newblock {\em Eur. Phys. J. C}, 81(3):203, 2021.

\bibitem{Nunez2}
H.~Hern\'andez, D.~Su\'arez-Urango, and L.A. N\'u\~nez.
\newblock {Acceptability Conditions and Relativistic Barotropic Equations of
  State}.
\newblock {\em Eur. Phys. J. C}, 81(3):241, 2021.

\bibitem{Nunez}
D.~Su\'arez-Urango, L.~A. N\'u\~nez, and H.~Hern\'andez.
\newblock {Relativistic Anisotropic Polytropic Spheres: Physical
  Acceptability}.
\newblock 1 2021.

\bibitem{Stuchlik}
J.~Novotn\'y, J.~Hlad\'\i{}k, and Z.~Stuchl\'\i{}k.
\newblock {Polytropic spheres containing regions of trapped null geodesics}.
\newblock {\em Phys. Rev. D}, 95(4):043009, 2017.

\bibitem{Stuchlik2}
C.~Posada, J.~Hlad\'\i{}k, and Z.~Stuchl\'\i{}k.
\newblock {Dynamical instability of polytropic spheres in spacetimes with a
  cosmological constant}.
\newblock {\em Phys. Rev. D}, 102(2):024056, 2020.

\bibitem{Stuchlik3}
J.~Hlad\'\i{}k, C.~Posada, and Z.~Stuchl\'\i{}k.
\newblock {Radial instability of trapping polytropic spheres}.
\newblock {\em Int. J. Mod. Phys. D}, 29(05):2050030, 2020.

\bibitem{Stuchlik4}
J.~Novotn\'y, Z.~Stuchl\'\i{}k, and J.~Hlad\'\i{}k.
\newblock {Polytropic spheres modelling dark matter haloes of dwarf galaxies}.
\newblock {\em Astron. Astrophys.}, 647:A29, 2021.

\bibitem{Chavanis2}
P.~H.~Chavanis,
Eur. Phys. J. Plus \textbf{129}, no.10, 222 (2014)
doi:10.1140/epjp/i2014-14222-0
[arXiv:1208.0801 [astro-ph.CO]].

\bibitem{Bowers}
R.~L. Bowers and E.P.T. Liang.
\newblock {Anisotropic Spheres in General Relativity}.
\newblock {\em Astrophys. J.}, 188:657--665, 1974.

\bibitem{Herrera10}
R.~Chan, L.~Herrera, and N.~O. Santos.
\newblock {Dynamical instability for radiating anisotropic collapse}.
\newblock {\em Monthly Notices of the Royal Astronomical Society},
  265(3):533--544, 12 1993.

\bibitem{Herrera2}
L.~Herrera and N.O. Santos.
\newblock Local anisotropy in self-gravitating systems.
\newblock {\em Physics Reports}, 286(2):53 -- 130, 1997.

\bibitem{Herrera11}
L.~Herrera, J.~Martin, and J.~Ospino.
\newblock {Anisotropic geodesic fluid spheres in general relativity}.
\newblock {\em J. Math. Phys.}, 43:4889--4897, 2002.

\bibitem{Herrera3}
L.~Herrera, A.~Di~Prisco, J.~Martin, J.~Ospino, N.~O. Santos, and O.~Troconis.
\newblock {Spherically symmetric dissipative anisotropic fluids: A General
  study}.
\newblock {\em Phys. Rev.}, D69:084026, 2004.

\bibitem{Rincon5}
F.~Tello-Ortiz, M.~Malaver, A.~Rinc\'on, and Y.~Gomez-Leyton.
\newblock {Relativistic anisotropic fluid spheres satisfying a non-linear
  equation of state}.
\newblock {\em Eur. Phys. J. C}, 80(5):371, 2020.

\bibitem{Rincon6}
G.~Panotopoulos, A.~Rinc\'on, and I.~Lopes.
\newblock {Interior solutions of relativistic stars with anisotropic matter in
  scale-dependent gravity}.
\newblock {\em Eur. Phys. J. C}, 81(1):63, 2021.

\bibitem{Rincon7}
I.~Lopes, G.~Panotopoulos, and A.~Rinc\'on.
\newblock {Anisotropic strange quark stars with a non-linear
  equation-of-state}.
\newblock {\em Eur. Phys. J. Plus}, 134(9):454, 2019.

\bibitem{Rincon8}
G.~Panotopoulos and .~Rinc\'on.
\newblock {Relativistic strange quark stars in Lovelock gravity}.
\newblock {\em Eur. Phys. J. Plus}, 134(9):472, 2019.

\bibitem{Herrera6}
L.~Herrera.
\newblock {New definition of complexity for self-gravitating fluid
  distributions: The spherically symmetric, static case}.
\newblock {\em Phys. Rev. D}, 97(4):044010, 2018.

\bibitem{Lopez3}
R.~G. Catal\'an, J.~Garay, and R.~L\'opez-Ruiz.
\newblock Features of the extension of a statistical measure of complexity to
  continuous systems.
\newblock {\em Phys. Rev. E}, 66:011102, Jul 2002.

\bibitem{Sanudo}
J.~Sanudo and A.~F. Pacheco.
\newblock {Complexity and white-dwarf structure}.
\newblock {\em Phys. Lett. A}, 373:807--810, 2009.

\bibitem{Chatzisavvas}
K.~Ch. Chatzisavvas, V.~P. Psonis, C.~P. Panos, and Ch.~C. Moustakidis.
\newblock {Complexity and neutron stars structure}.
\newblock {\em Phys. Lett. A}, 373:3901--3909, 2009.

\bibitem{DEAVELLAR}
M.G.B. {de Avellar} and J.E. Horvath.
\newblock Entropy, complexity and disequilibrium in compact stars.
\newblock {\em Physics Letters A}, 376(12):1085--1089, 2012.

\bibitem{DEAVELLAR2}
M.~G.~B. {de Avellar} and J.~E. {Horvath}.
\newblock {Entropy, Disequilibrium and Complexity in Compact Stars: An
  information theory approach to understand their Composition}.
\newblock {\em arXiv e-prints}, page arXiv:1308.1033, August 2013.

\bibitem{DEAVELLAR3}
R.~A. de~Souza, M.~G. de~Avellar, and J.~E. Horvath.
\newblock {Statistical measure of complexity in compact stars with global
  charge neutrality}.
\newblock In {\em {Compact Stars in the QCD Phase Diagram III}}, 8 2013.

\bibitem{DEAVELLAR4}
M.~G.~B. de~Avellar, R.~A. de~Souza, J.~E. Horvath, and D.~M. Paret.
\newblock {Information theoretical methods as discerning quantifiers of the
  equations of state of neutron stars}.
\newblock {\em Phys. Lett. A}, 378:3481--3487, 2014.

\bibitem{Herrera7}
L.~Herrera, A.~Di~Prisco, and J.~Carot.
\newblock {Complexity of the Bondi metric}.
\newblock {\em Phys. Rev. D}, 99(12):124028, 2019.

\bibitem{Herrera8}
L.~Herrera, A.~Di~Prisco, and J.~Ospino.
\newblock {Definition of complexity for dynamical spherically symmetric
  dissipative self-gravitating fluid distributions}.
\newblock {\em Phys. Rev. D}, 98(10):104059, 2018.

\bibitem{Abbas}
G.~Abbas and H.~Nazar.
\newblock {Complexity Factor For Static Anisotropic Self-Gravitating Source in
  $f(R)$ Gravity}.
\newblock {\em Eur. Phys. J. C}, 78(6):510, 2018.

\bibitem{Sharif10}
M.~Sharif and I.~I. Butt.
\newblock {Complexity Factor for Charged Spherical System}.
\newblock {\em Eur. Phys. J. C}, 78(8):688, 2018.

\bibitem{Sharif:2022wnl}
M.~Sharif and A.~Majid,
\newblock {Isotropization and complexity of decoupled solutions in self-interacting Brans\textendash{}Dicke gravity}.
\newblock {Eur. Phys. J. Plus}, 137(1):114, 2022.

\bibitem{Sharif:2021gsl}
M.~Sharif and K.~Hassan,
\newblock {Complexity of dynamical cylindrical system in $f(G, T)$ gravity}.
\newblock {Mod. Phys. Lett. A}, 37(05): 2250027, 2022.

\bibitem{Zubair50}
M.~Zubair and H.~Azmat,
\newblock {Complexity analysis of cylindrically symmetric self-gravitating dynamical system in f(R,T) theory of gravity}.
\newblock {\em Phys. Dark Univ.}, 28:100531, 2020.

\bibitem{Zubair51}
M.~Zubair and H.~Azmat,
\newblock {Complexity analysis of dynamical spherically-symmetric dissipative self-gravitating objects in modified gravity}.
\newblock {\em nt. J. Mod. Phys.}, 29(02):2050014, 2020.

\bibitem{Yousaf}
Z.~Yousaf, M.~Z. Bhatti, S.~Khan, and P.~K. Sahoo.
\newblock
  {f(G,T\ensuremath{\alpha}\ensuremath{\beta}T\ensuremath{\alpha}\ensuremath{\beta})
  theory and complex cosmological structures}.
\newblock {\em Phys. Dark Univ.}, 36:101015, 2022.

\bibitem{Yousaf2}
Z.~Yousaf.
\newblock {Definition of complexity factor for self-gravitating systems in
  Palatini $f(R)$ gravity}.
\newblock {\em Phys. Scripta}, 95(7):075307, 2020.

\bibitem{Yousaf3}
Z.~Yousaf, M.~Y. Khlopov, M.~Z. Bhatti, and T.~Naseer.
\newblock {Influence of Modification of Gravity on the Complexity Factor of
  Static Spherical Structures}.
\newblock {\em Mon. Not. Roy. Astron. Soc.}, 495(4):4334--4346, 2020.

\bibitem{Yousaf4}
Z.~Yousaf, M.~Z. Bhatti, and S.~Khan.
\newblock {Non-static charged complex structures in $f({\mathbb {G}}, {\mathbf
  {T}}^2)$ gravity}.
\newblock {\em Eur. Phys. J. Plus}, 137(3):322, 2022.

\bibitem{Yousaf5}
M.~Z. Bhatti, M.~Y. Khlopov, Z.~Yousaf, and S.~Khan.
\newblock {Electromagnetic field and complexity of relativistic fluids in f (G)
  gravity}.
\newblock {\em Mon. Not. Roy. Astron. Soc.}, 506(3):4543--4560, 2021.

\bibitem{Yousaf40}
Z.~Yousaf, M.~Z. Bhatti, and T.~Naseer.
\newblock {Measure of complexity for dynamical self-gravitating structures}.
\newblock {\em Int. J. Mod. Phys. D},29(09):2050061, 2020.

\bibitem{Maurya:2022cyv}
S.~K.~Maurya, A.~Errehymy, R.~Nag and M.~Daoud,
\newblock {Role of Complexity on Self-gravitating Compact Star by Gravitational Decoupling}.
\newblock {\em Fortsch. Phys.} 70 (5): 2200041, 2022.

\bibitem{Maurya:2022cyv2}
S.~K.~Maurya, M.~Govender, S.~Kaur and R.~Nag,
\newblock {Isotropization of embedding class I spacetime and anisotropic system generated by complexity factor in the framework of gravitational decoupling}.
\newblock {\em Eur. Phys. J. C} 82 (2): 100, 2022.

\bibitem{Contreras:2022vec2}
E.~Contreras and Z.~Stuchlik,
\newblock {A simple protocol to construct solutions with vanishing complexity by Gravitational Decoupling}.
\newblock {\em Eur. Phys. J. C} 82 (8):706, 2022.

\bibitem{Ovalle}
J.~Ovalle.
\newblock {Decoupling gravitational sources in general relativity: from perfect
  to anisotropic fluids}.
\newblock {\em Phys. Rev.}, D95(10):104019, 2017.

\bibitem{Ovalle1}
J.~Ovalle.
\newblock {Searching exact solutions for compact stars in braneworld: A
  Conjecture}.
\newblock {\em Mod. Phys. Lett.}, A23:3247--3263, 2008.

\bibitem{Ovalle15}
J.~Ovalle and R.~Casadio.
\newblock {\em {Beyond Einstein Gravity}}.
\newblock SpringerBriefs in Physics. Springer, 1 2020.

\bibitem{Ovalle2}
J~Ovalle.
\newblock {Braneworld Stars: Anisotropy Minimally Projected Onto the Brane}.
\newblock In {\em {9th Asia-Pacific International Conference on Gravitation and
  Astrophysics (ICGA 9) Wuhan, China, June 28-July 2, 2009}}, pages 173--182,
  2009.

\bibitem{Ovalle16}
R.~Casadio and J.~Ovalle.
\newblock {Brane-world stars from minimal geometric deformation, and black
  holes}.
\newblock {\em Gen. Rel. Grav.}, 46:1669, 2014.

\bibitem{Ovalle8}
J.~Ovalle, F.~Linares, A.~Pasqua, and A.~Sotomayor.
\newblock {The role of exterior Weyl fluids on compact stellar structures in
  Randall-Sundrum gravity}.
\newblock {\em Class. Quant. Grav.}, 30:175019, 2013.

\bibitem{Ovalle9}
R.~Casadio, J.~Ovalle, and R.~da~Rocha.
\newblock {Black Strings from Minimal Geometric Deformation in a Variable
  Tension Brane-World}.
\newblock {\em Class. Quant. Grav.}, 31:045016, 2014.

\bibitem{Ovalle7}
J~Ovalle and F~Linares.
\newblock {Tolman IV solution in the Randall-Sundrum Braneworld}.
\newblock {\em Phys. Rev.}, D88(10):104026, 2013.

\bibitem{Ovalle10}
J.~Ovalle, L.A. Gergely, and R.~Casadio.
\newblock {Brane-world stars with a solid crust and vacuum exterior}.
\newblock {\em Class. Quant. Grav.}, 32:045015, 2015.

\bibitem{Ovalle11}
R.~Casadio, J.~Ovalle, and R.~da~Rocha.
\newblock {Classical Tests of General Relativity: Brane-World Sun from Minimal
  Geometric Deformation}.
\newblock {\em EPL}, 110(4):40003, 2015.

\bibitem{Our}
C.~Las~Heras and P.~Leon.
\newblock {Using MGD gravitational decoupling to extend the isotropic solutions
  of Einstein equations to the anisotropical domain}.
\newblock {\em Fortsch. Phys.}, 66(7):1800036, 2018.

\bibitem{Estrada1}
M.~Estrada and F.~Tello-Ortiz.
\newblock {A new family of analytical anisotropic solutions by gravitational
  decoupling}.
\newblock {\em Eur. Phys. J. Plus}, 133(11):453, 2018.

\bibitem{Gabbanelli}
L.~Gabbanelli, A.~Rincón, and C.~Rubio.
\newblock {Gravitational decoupled anisotropies in compact stars}.
\newblock {\em Eur. Phys. J.}, C78(5):370, 2018.

\bibitem{Morales1}
E.~Morales and F.~Tello-Ortiz.
\newblock {Compact Anisotropic Models in General Relativity by Gravitational
  Decoupling}.
\newblock {\em Eur. Phys. J.}, C78(10):841, 2018.

\bibitem{Morales2}
E.~Morales and F.~Tello-Ortiz.
\newblock {Charged anisotropic compact objects by gravitational decoupling}.
\newblock {\em Eur. Phys. J.}, C78(8):618, 2018.

\bibitem{Tello2}
F.~Tello-Ortiz, S.K. Maurya, A.~Errehymy, K.N. Singh, and M.~Daoud.
\newblock {Anisotropic relativistic fluid spheres: an embedding class I
  approach}.
\newblock {\em Eur. Phys. J. C}, 79(11):885, 2019.

\bibitem{Contreras7}
V.A. Torres-S\'anchez and E.~Contreras.
\newblock {Anisotropic neutron stars by gravitational decoupling}.
\newblock {\em Eur. Phys. J. C}, 79(10):829, 2019.

\bibitem{Ovalle13}
J.~Ovalle, R.~Casadio, R.~da~Rocha, A.~Sotomayor, and Z.~Stuchlik.
\newblock {Einstein-Klein-Gordon system by gravitational decoupling}.
\newblock {\em EPL}, 124(2):20004, 2018.

\bibitem{Sharif3}
M.~Sharif and A.~Waseem.
\newblock Anisotropic spherical solutions by gravitational decoupling in $f(r)$
  gravity.
\newblock {\em Annals of Physics}, 405:14 -- 28, 2019.

\bibitem{Sharif4}
M.~Sharif and S.~Saba.
\newblock Extended gravitational decoupling approach in $f(\mathcal {G})$
  gravity.
\newblock {\em International Journal of Modern Physics D}, 29(06):2050041,
  2020.

\bibitem{Sharif5}
M.~Sharif and S.~Saba.
\newblock {Gravitational decoupled anisotropic solutions in $f(\mathcal {G})$
  gravity}.
\newblock {\em Eur. Phys. J. C}, 78(11):921, 2018.

\bibitem{Tello4}
S.~K. Maurya, A.~Errehymy, K.N. Singh, F.~Tello-Ortiz, and M.~Daoud.
\newblock {Gravitational decoupling minimal geometric deformation model in
  modified $f(R,T)$ gravity theory}.
\newblock {\em Phys. Dark Univ.}, 30:100640, 2020.

\bibitem{Estrada3}
M.~Estrada.
\newblock {A way of decoupling gravitational sources in pure Lovelock gravity}.
\newblock {\em Eur. Phys. J. C}, 79(11):918, 2019.
\newblock [Erratum: Eur.Phys.J.C 80, 590 (2020)].

\bibitem{Leon}
P.~Le\'on and A.~Sotomayor.
\newblock {Braneworld Gravity under gravitational decoupling}.
\newblock {\em Fortsch. Phys.}, 67(12):1900077, 2019.

\bibitem{Leon2}
P.~León and A.~Sotomayor.
\newblock Braneworld-klein-gordon system in the framework of gravitational
  decoupling.
\newblock {\em Fortschritte der Physik}, 69(10):2100017, 2021.

\bibitem{Ovalle12}
J.~Ovalle.
\newblock {Decoupling gravitational sources in general relativity: The extended
  case}.
\newblock {\em Phys. Lett. B}, 788:213--218, 2019.

\bibitem{Contreras13}
E.~Contreras, J.~Ovalle, and R.~Casadio.
\newblock {Gravitational decoupling for axially symmetric systems and rotating
  black holes}.
\newblock 1 2021.

\bibitem{Ovalle6}
R.~Casadio and J.~Ovalle.
\newblock {Brane-world stars and (microscopic) black holes}.
\newblock {\em Phys. Lett.}, B715:251--255, 2012.

\bibitem{Ovalle17}
J.~Ovalle, R.~Casadio, R.~da~Rocha, and A.~Sotomayor.
\newblock {Anisotropic solutions by gravitational decoupling}.
\newblock {\em Eur. Phys. J. C}, 78(2):122, 2018.

\bibitem{Ovalle18}
L.~Gabbanelli, J.~Ovalle, A.~Sotomayor, Z.~Stuchlik, and R.~Casadio.
\newblock {A causal Schwarzschild-de Sitter interior solution by gravitational
  decoupling}.
\newblock {\em Eur. Phys. J. C}, 79(6):486, 2019.

\bibitem{Ovalle19}
R.~Casadio, E.~Contreras, J.~Ovalle, A.~Sotomayor, and Z.~Stuchlick.
\newblock {Isotropization and change of complexity by gravitational
  decoupling}.
\newblock {\em Eur. Phys. J. C}, 79(10):826, 2019.

\bibitem{Ovalle20}
J~Ovalle, C~Posada, and Z~Stuchl{\'{\i}}k.
\newblock 36(20):205010, sep 2019.

\bibitem{Abellan}
G.~Abell\'an, V.~A. Torres-S\'anchez, E.~Fuenmayor, and E.~Contreras.
\newblock{Regularity condition on the anisotropy induced by gravitational
  decoupling in the framework of MGD}.
\newblock {\em Eur. Phys. J. C}, 80(2):177, 2020.

\bibitem{Abellan4}
G.~Abell\'an, A.~Rincon, E.~Fuenmayor, and E.~Contreras.
\newblock{Beyond classical anisotropy and a new look to relativistic stars: a
  gravitational decoupling approach}.
  \newblock{arXiv:2001.07961 [gr-qc]}
\newblock 2020.

\bibitem{Sharif}
M.~Sharif and Q.~Ama-Tul-Mughani.
\newblock {Anisotropic Spherical Solutions through Extended Gravitational
  Decoupling Approach}.
\newblock {\em Annals Phys.}, 415:168122, 2020.

\bibitem{Sharif2}
M.~Sharif and A.~Majid.
\newblock {Extended gravitational decoupled solutions in self-interacting
  Brans-Dicke theory}.
\newblock {\em Phys. Dark Univ.}, 30:100610, 2020.

\bibitem{Sharif8}
M.~Sharif and A.~Majid.
\newblock {Decoupled anisotropic spheres in self-interacting Brans-Dicke
  gravity}.
\newblock {\em Chin. J. Phys.}, 68:406--418, 2020.

\bibitem{Sharif9}
M.~Sharif and A.~Majid.
\newblock {Extended gravitational decoupled solutions in self-interacting
  Brans-Dicke theory}.
\newblock {\em Phys. Dark Univ.}, 30:100610, 2020.

\bibitem{Cavalcanti}
R.~T. Cavalcanti, A.~Goncalves da~Silva, and R.~da~Rocha.
\newblock {Strong deflection limit lensing effects in the minimal geometric
  deformation and Casadio-Fabbri-Mazzacurati solutions}.
\newblock {\em Class. Quant. Grav.}, 33(21):215007, 2016.

\bibitem{Darocha1}
R.~da~Rocha.
\newblock {Dark SU(N) glueball stars on fluid branes}.
\newblock {\em Phys. Rev.}, D95(12):124017, 2017.

\bibitem{Darocha2}
R.~da~Rocha.
\newblock {Black hole acoustics in the minimal geometric deformation of a de
  Laval nozzle}.
\newblock {\em Eur. Phys. J.}, C77(5):355, 2017.

\bibitem{Darocha3}
A.~Fernandes-Silva and R.~da~Rocha.
\newblock {Gregory-Laflamme analysis of MGD black strings}.
\newblock {\em Eur. Phys. J.}, C78(3):271, 2018.

\bibitem{Darocha4}
A.~Fernandes-Silva, A.~J. Ferreira-Martins, and R.~Da~Rocha.
\newblock {The extended minimal geometric deformation of SU($N$) dark glueball
  condensates}.
\newblock {\em Eur. Phys. J. C}, 78(8):631, 2018.

\bibitem{Darocha5}
R.~Da~Rocha and A.~Tomaz.
\newblock {Holographic entanglement entropy under the minimal geometric
  deformation and extensions}.
\newblock {\em Eur. Phys. J. C}, 79(12):1035, 2019.

\bibitem{Darocha6}
R~da~Rocha.
\newblock {MGD Dirac stars}.
\newblock {\em Symmetry}, 12(4):508, 2020.

\bibitem{Darocha7}
R.~da~Rocha.
\newblock {Minimal geometric deformation of Yang-Mills-Dirac stellar
  configurations}.
\newblock {\em Phys. Rev. D}, 102(2):024011, 2020.

\bibitem{Darocha8}
R.~da~Rocha and A.~Tomaz.
\newblock {MGD-decoupled black holes, anisotropic fluids and holographic
  entanglement entropy}.
\newblock {\em Eur. Phys. J. C}, 80(9):857, 2020.

\bibitem{Darocha9}
P.~Meert and R.~da~Rocha.
\newblock {Probing the minimal geometric deformation with trace and Weyl
  anomalies}.
\newblock {\em Nucl. Phys. B}, 967:115420, 2021.

\bibitem{Casadio2}
R.~Casadio, P.~Nicolini, and R.~da~Rocha.
\newblock {Generalised uncertainty principle Hawking fermions from minimally
  geometric deformed black holes}.
\newblock {\em Class. Quant. Grav.}, 35(18):185001, 2018.

\bibitem{Contreras}
E.~Contreras and P.~Bargue\~no.
\newblock {Extended gravitational decoupling in 2 + 1 dimensional space-times}.
\newblock {\em Class. Quant. Grav.}, 36(21):215009, 2019.

\bibitem{Contreras2}
E.~Contreras, Á. Rincón, and P.~Bargueño.
\newblock {A general interior anisotropic solution for a BTZ vacuum in the
  context of the Minimal Geometric Deformation decoupling approach}.
\newblock {\em Eur. Phys. J.}, C79(3):216, 2019.

\bibitem{Contreras9}
A.~Rinc\'on, E.~Contreras, F.~Tello-Ortiz, P.~Bargue\~no, and G.~Abell\'an.
\newblock {Anisotropic 2+1 dimensional black holes by gravitational
  decoupling}.
\newblock {\em Eur. Phys. J. C}, 80(6):490, 2020.

\bibitem{Contreras5}
E.~Contreras.
\newblock {Minimal Geometric Deformation: the inverse problem}.
\newblock {\em Eur. Phys. J.}, C78(8):678, 2018.

\bibitem{Contreras14}
E.~Contreras, F.~Tello-Ortiz, and S.~K. Maurya.
\newblock {Regular decoupling sector and exterior solutions in the context of
  MGD}.
\newblock {\em Class. Quant. Grav.}, 37(15):155002, 2020.

\bibitem{Contreras15}
C.~Arias, F.~Tello-Ortiz, and E.~Contreras.
\newblock {Extra packing of mass of anisotropic interiors induced by MGD}.
\newblock {\em Eur. Phys. J. C}, 80(5):463, 2020.

\bibitem{Rincon}
G.~Panotopoulos and A.~Rinc\'on.
\newblock {Minimal Geometric Deformation in a cloud of strings}.
\newblock {\em Eur. Phys. J. C}, 78(10):851, 2018.

\bibitem{Rincon2}
L.~Gabbanelli, A.~Rinc\'on, and C~Rubio.
\newblock {Gravitational decoupled anisotropies in compact stars}.
\newblock {\em Eur. Phys. J. C}, 78(5):370, 2018.

\bibitem{Rincon3}
A.~Rinc\'on, L.~Gabbanelli, E.~Contreras, and F.~Tello-Ortiz.
\newblock {Minimal geometric deformation in a Reissner\textendash{}Nordstr\"om
  background}.
\newblock {\em Eur. Phys. J. C}, 79(10):873, 2019.

\bibitem{Rincon4}
F.~Tello-Ortiz, A.~Rinc\'on, P.~Bhar, and Y.~Gomez-Leyton.
\newblock {Durgapal IV model in light of the minimal geometric deformation
  approach}.
\newblock {\em Chin. Phys. C}, 44:105102, 2020.

\bibitem{Tello6}
S.~K. Maurya and F.~Tello-Ortiz.
\newblock {Generalized relativistic anisotropic compact star models by
  gravitational decoupling}.
\newblock {\em Eur. Phys. J. C}, 79(1):85, 2019.

\bibitem{Tello7}
S.~K. Maurya and F.~Tello-Ortiz.
\newblock {Charged anisotropic compact star in $f(R,T)$ gravity: A minimal
  geometric deformation gravitational decoupling approach}.
\newblock {\em Phys. Dark Univ.}, 27:100442, 2020.

\bibitem{Hensh}
S.~Hensh and Z.~Stuchl\'i~k.
\newblock {Anisotropic Tolman VII solution by gravitational decoupling}.
\newblock {\em Eur. Phys. J. C}, 79(10):834, 2019.

\bibitem{Maurya2}
K.N. Singh, S.K. Maurya, M.K. Jasim, and F.~Rahaman.
\newblock {Minimally deformed anisotropic model of class one space-time by
  gravitational decoupling}.
\newblock {\em Eur. Phys. J. C}, 79(10):851, 2019.

\bibitem{Maurya3}
S.~K. {Maurya}.
\newblock {A completely deformed anisotropic class one solution for charged
  compact star: a gravitational decoupling approach}.
\newblock {\em European Physical Journal C}, 79(11):958, November 2019.

\bibitem{Maurya4}
S.~K. Maurya.
\newblock {Extended gravitational decoupling (GD) solution for charged compact
  star model}.
\newblock {\em Eur. Phys. J. C}, 80(5):429, 2020.

\bibitem{Maurya5}
S.~K. Maurya.
\newblock {Non-singular solution for anisotropic model by gravitational
  decoupling in the framework of complete geometric deformation (CGD)}.
\newblock {\em Eur. Phys. J. C}, 80(5):448, 2020.

\bibitem{Zubair}
M.~Zubair and H.~Azmat.
\newblock {Anisotropic Tolman V Solution by Minimal Gravitational Decoupling
  Approach}.
\newblock {\em Annals Phys.}, 420:168248, 2020.

\bibitem{Nariai}
H.~{Nariai}.
\newblock {On some static solutions of Einstein's gravitational field equations
  in a spherically symmetric case}.
\newblock {\em Sci.~Rep.~Tohoku Univ.~Eighth Ser.}, 34, 1950.

\bibitem{Our2}
C.~Las~Heras and P.~Le\'on.
\newblock {New algorithms to obtain analytical solutions of Einstein's
  equations in isotropic coordinates}.
\newblock {\em Eur. Phys. J. C}, 79(12):990, 2019.

\bibitem{Our3}
C.~Las~Heras and P.~Leon.
\newblock {New interpretation of the extended geometric deformation in
  isotropic coordinates}.
\newblock {\em Eur. Phys. J. Plus}, 136(8):828, 2021.

\bibitem{Contreras4}
E.~Contreras and P.~Bargueño.
\newblock {Minimal Geometric Deformation in asymptotically (A-)dS space-times
  and the isotropic sector for a polytropic black hole}.
\newblock {\em Eur. Phys. J.}, C78(12):985, 2018.

\bibitem{Contreras16}
E.~Contreras and P.~Bargue\~no.
\newblock {Extended gravitational decoupling in 2 + 1 dimensional space-times}.
\newblock {\em Class. Quant. Grav.}, 36(21):215009, 2019.

\bibitem{Maurya7}
S.~K. Maurya, M.~Govender, K.N. Singh, and R.~Nag.
\newblock {Gravitationally decoupled anisotropic solution using polytropic EoS
  in the framework of 5D Einstein\textendash{}Gauss\textendash{}Bonnet
  Gravity}.
\newblock {\em Eur. Phys. J. C}, 82(1):49, 2022.

\bibitem{Contreras20}
J.~Ovalle, E.~Contreras, and Z.~Stuchlik.
\newblock {Energy exchange between relativistic fluids: the polytropic case}.
\newblock {\em Eur. Phys. J. C}, 82(3):211, 2022.

\bibitem{Contreras8}
R.~Casadio, E.~Contreras, J.~Ovalle, A.~Sotomayor, and Z.~Stuchlick.
\newblock {Isotropization and change of complexity by gravitational
  decoupling}.
\newblock {\em Eur. Phys. J. C}, 79(10):826, 2019.

\bibitem{Contreras18}
E.~Contreras and E.~Fuenmayor.
\newblock {Gravitational cracking and complexity in the framework of
  gravitational decoupling}.
\newblock {\em Phys. Rev. D}, 103(12):124065, 2021.

\bibitem{Contreras19}
M.~Carrasco-Hidalgo and E.~Contreras.
\newblock {Ultracompact stars with polynomial complexity by gravitational
  decoupling}.
\newblock {\em Eur. Phys. J. C}, 81(8):757, 2021.

\bibitem{Maurya6}
S.~K. Maurya and R.~Nag.
\newblock {Role of gravitational decoupling on isotropization and complexity of
  self-gravitating system under complete geometric deformation approach}.
\newblock {\em Eur. Phys. J. C}, 82(1):48, 2022.

\bibitem{Schwarzschild2}
Deutsche~Akademie der Wissenschaften~zu Berlin.
\newblock {\em Sitzungsberichte der Königlich Preussischen Akademie der
  Wissenschaften zu Berlin}, volume Jan-Juni 1916.
\newblock Berlin, Deutsche Akademie der Wissenschaften zu Berlin, 1882-1918,
  1916.
\newblock https://www.biodiversitylibrary.org/bibliography/42231.

\bibitem{Our4}
C.~Las~Heras and P.~Leon.
\newblock {(To appear)}.
\newblock 2022.

\bibitem{Boonserm}
P.~Boonserm, M.~Visser, S.~Weinfurtner.
\newblock {Generating perfect fluid spheres in general relativity}
\newblock {\em Phys. Rev. D}, 71:124037, 2005.

\bibitem{Andrade}
J.~Andrade.
\newblock {Stellar solutions with zero complexity obtained through a temporal
  metric deformation}.
\newblock {\em Eur. Phys. J. C}, 82(3):266, 2022.

\end{thebibliography}
\end{document}